# Dual Micro-Ring Resonators with Angular GST Modulation: Enabling Ultra-Fast Nonlinear Activation for Neuromorphic Photonics

Hossein Karimkhani[1], Yaser M. Banad[1], and Sarah Sharif[1,2*]

[1]School of Electrical and Computer Engineering, University of Oklahoma, Norman, OK, USA, 73019

Center for Quantum Research and Technology, University of Oklahoma, Norman, OK, USA, 73019

[*]E-mail: s.sh@ou.edu

**Abstract:** Photonic technologies are emerging as powerful enablers for neuromorphic computing by delivering ultrafast and energy-efficient neural functionalities. In this work, we propose and demonstrate a novel all-optical dual micro-ring resonator architecture incorporating the phase-change material $Ge_2Sb_2Te_5$ (GST) to implement highly precise nonlinear activation functions (NLAFs). Our approach introduces angular positioning of GST segments within the rings, enabling fine-grained control over optical transmission dynamics. Through a systematic evaluation of sixteen distinct phase configurations, we identify an optimal GST placement—180° in the first ring and 90° in the second—that achieves ultra-narrowband transmission with a full width at half maximum (FWHM) of just 0.47 nm. This dual-ring configuration provides two distinct resonant wavelengths, facilitating enhanced nonlinear modulation and multi-level optical signal processing that closely mimics biological neuron behavior. Notably, the device achieves high contrast transmission (0 to 0.85) across a 4 nm spectral window while operating at significantly reduced temperatures (~100 °C), outperforming traditional GST-based designs. Furthermore, the dual-ring architecture enables independent optimization of spectral selectivity and switching contrast—capabilities previously unattainable with single-ring structures. These results establish a promising pathway toward scalable, high-speed neuromorphic photonic systems, offering both the precision and switching speed required for practical on-chip neural processing.

## 1. Introduction

   The rapid expansion of information technology, particularly in computing, has led to a significant increase in global data traffic [1,2]. To meet this demand, artificial intelligence (AI) technologies, which leverage neural networks, are advancing to deliver faster processing. However, traditional computing architectures are increasingly constrained by energy efficiency and scalability limitations [3].
   Initial research focused on digital electronic systems, traditional computing paradigms that process information using binary signals (0s and 1s) within electronic circuits [4,5]. These systems exhibit inefficiencies and limited computational speeds, prompting researchers to enhance them through neuromorphic electronics designed to emulate neural networks [6]. While these systems offer improved efficiency and computational speed relative to traditional digital electronics, they still face challenges in achieving optimal performance for high-speed applications [7].
   Electronic neuromorphic computing, which utilizes CMOS-based architectures to emulate neural behaviors, has been a significant step forward [8,9]. It provides faster, more energy-efficient solutions than traditional von Neumann architectures. However, as data processing demands continue to grow, electronic systems face physical limits related to heat dissipation, latency, and energy consumption, particularly when scaling to large, complex networks [2,10,11].
   To address these challenges, there has been a notable surge in advancements within brain-inspired computing. In this context, the human brain serves as a reference point for achieving high efficiency with low energy consumption [12,13]. There is an escalating interest in developing efficient photonic hardware systems that focus on the brain's functional units, such as neurons. Neurons are fundamental components of neural networks, and the optical neurons play a vital role in optical neural networks and neuromorphic photonics [14]. These systems employ Nonlinear Activation Functions (NLAFs) to perform nonlinear computations [15]. In neuromorphic photonic devices, NLAFs are crucial for completing the mapping processes within optical networks, converting linear outputs into nonlinear results essential for system processing [16]. Selecting appropriate NLAFs based on specific tasks is critical for optimal performance. Current NLAF devices require further optimization, as they have not been specifically designed for their intended applications, leading to higher energy demands for converting electrical signals to optical signals and introducing complexity that affects efficiency [17].

Despite various attempts to implement spiking neuron models, such as Hodgkin-Huxley and Leaky-Integrate-Fire, on CMOS platforms [18,19], these hardware implementations fall short regarding energy and area efficiency compared to the human brain [20,21]. In response, silicon photonic devices emerged, utilizing photons for computation and providing even greater efficiency and speed [7]; however, these rates remain insufficient for demanding applications [22–24]. Consequently, researchers have developed neuromorphic photonic devices to achieve enhanced efficiency and high-speed computational capabilities [25]. In recent years, substantial progress has been made in this area.

Spiking Neural Networks (SNNs) leverage biologically plausible neurons to facilitate faster inference, lower energy consumption, and event-driven processing, bridging the gap between artificial neural networks (ANNs) and biological neural networks (BNNs) [1,26,27]. Implementing SNNs in future neuromorphic hardware necessitates neuron-like hardware encoders to convert inputs into spike trains [28–30]. Recent advancements suggest that SNNs present a promising computational model, enhancing energy efficiency and inference speed through event-driven computation [2,31].

Phase Change Materials (PCMs) have emerged as a potential solution to the challenges of implementing spiking neural networks [32,33]. These materials can mimic neural spiking activity, allowing a single device to function as a single neural network [14]. PCMs exhibit extraordinary properties during phase transitions between amorphous and crystalline states, involving the absorption or release of substantial energy [2].

The reversible phase-switching of materials between these states has been explored since the 1960s, initially in studies on colonic threshold switching in disordered structures [34]. For most applications, an optimal PCM should exhibit high-speed, low-power switching capabilities, durability over numerous switching cycles, long-term thermal stability in the amorphous state, and significant optical or electrical contrast between the two phases [35].

Among potential PCMs, chalcogenide glasses based on germanium (Ge), antimony (Sb), and tellurium (Te) alloys collectively known as GST, with $Ge_2Sb_2Te_5$ as a notable example—excel in meeting these criteria [35–37]. GST material exhibits a rapid phase transition from amorphous to crystalline states, enabling efficient data storage and retrieval [28,38]. The nonlinear optical properties of GST make it highly suitable for photonic applications [39,40].

GST can be quickly excited by output laser pulses and temperature variations, demonstrating faster times compared to CMOS devices [32,41]. Its distinct electrical and optical properties in both amorphous and crystalline states make it useful in optical switches and resonators [28,40]. The potential for high-speed information processing and transmission through PCMs in photonic structures has positioned materials like GST as key candidates for photonic neuromorphic devices [42,43]. Recent advances have demonstrated various GST-based structures and optical waveguides capable of simulating the synaptic weight update mechanism in SNNs. Previous studies on SNNs in photonic devices often relied on converting electrical signals to optical signals, using lasers to emulate spiking neuron activity [2,14,28,44]. The integration of GST in photonic devices presents a promising avenue for developing efficient, high-speed neuromorphic systems.

To understand the context and significance of GST-based neuromorphic photonic systems, it is important to trace the evolution of optical neural networks and ring resonator technology. The concept of optical neural networks was first introduced in 1985 [45], and several studies have used free space optics and fiber components to build such networks [46]. However, these systems have remained confined to laboratory demonstrations due to the limited scalability strategies and immature fiber-optic and photonic technologies. Since then, the technology has progressed significantly, and neuroscience has also advanced, leading to an evolution in photonic neuromorphic research.

In 2009, the first successful demonstration of photonic spike processing took place [47]. In 2017, researchers proposed a fully integrated all-photonic synapse based on PCMs and implemented it via a photonic integrated-circuit approach [40]. This device closely resembles the neural synapse [28,48] at the physical level and can achieve synaptic plasticity compatible with the Spike Timing Dependent Plasticity (STDP) rule [6,49]. Following this work, another research group proposed an all-photonic integrate-and-fire spiking neuron that has the potential to be integrated into a spiking neural network [28]. Recently, an all-optical neuron system capable of supervised and unsupervised learning has been developed. It implements a simplified STDP rule through ring resonators and PCM, making it a promising avenue for future advancements in artificial intelligence [32]. Optical ring resonators are innovative optical devices developed for sensors [50], modulators [51], absorbers [52], and switches [53]. One of the main advantages of micro-ring resonators (MRR) is their high quality (Q) factors and small footprints, which allow them to operate across a wide range of wavelengths. Whispering Gallery Mode (WGM) ring resonators are optical structures with exceptional optical properties. They operate based on the principle of total internal reflection, allowing light waves to circulate along the curved boundaries. As a result, closed circular beams with high-quality factors are generated, facilitating efficient energy storage [54]. In 2018, a neuromorphic photonics system was proposed based on a ring resonator with a double rectangular waveguide. In the proposed device, the phase change material is GST and is directly placed on top of the ring resonator. The proposed structure consists of the through and drop ports. The GST on top of the ring can control the input light through the ports by changing the state of the material. The structure of this study contains a single ring resonator and a single transmission peak. They demonstrate that with increasing amorphization degree in the through port, transmission decreases, and in the drop port increases [28]. In 2022, a structure based on a series of ring resonators and waveguides purposed and GST layer with 20 nm height is placed on 330 nm ALN layer, which is placed on top of the ring resonator. The radius of the ring is 100 µm, and operating at 1500 nm telecommunication wavelength. The structure comprises a rectangular waveguide with a single input and output port. The proposed structure contains a pump with 800 nm wavelength and 35 fs duration, operating as a secondary source. This pump changes the reflectance and transmission rate by changing the temperature of the GST material. However, the proposed structure illustrates an ultra-narrow band transmission, but the footprint of the structure is large [2]. In 2023, another ring resonator with double rectangular waveguides and with embedded GST layer inside the ring was proposed. The radius of the ring is 5 µm, and the transmission line in through and drop ports is investigated. The transmission of the proposed structure is in the telecommunication wavelength, but it is not a narrow band [55].

Building on these advances, in this work we demonstrate a novel all-optical double micro ring resonator switch incorporating GST for neuromorphic computing applications. The proposed structure contains double-ring resonators based on all-optical NLAF. Our design integrates GST within silicon-based photonic components to achieve precise, temperature-dependent switching control. Through comprehensive analysis of multiple phase configurations and thermal responses, we show that this architecture enables high-speed switching while maintaining excellent transmission characteristics.

In this work, we demonstrate the first all-optical double micro-ring resonator based on GST material. This study simulates the effect of temperature-induced refractive index changes in GST materials, similar to experimental conditions involving laser stimulation. Our device, which has an angular positioning of GST segments within the rings, enables unprecedented control over optical transmission characteristics. Through systematic investigation of sixteen distinct phase configurations, we identify optimal GST positioning (180º in the first ring, 90º in the second) that achieves ultra-narrow band transmission with 0.47 nm full width at half maximum. The dual-ring design enables two distinct resonant wavelengths, providing enhanced nonlinear control and supporting multi-level optical processing. The dual-ring architecture enables independent optimization of spectral selectivity and switching contrast - a capability previously unattainable in single-ring designs. In this study, we decreased the radius of the ring to 5 µm to be able to decrease the footprint of the proposed device, however the radius of the ring resonator in the previous works considered between 6 µm to 100 µm [2,14,56].

## 2. Results and Analysis

### 2.1 Architecture and Concepts

Our neuromorphic photonic device implements a nonlinear activation function through a novel double micro-ring resonator architecture with embedded phase-change material. Fig. 1 illustrates the device's design principles and demonstrates its structural analogy to biological neural processing. The device architecture consists of a Si-based rectangular waveguide coupled to two ring resonators with integrated GST segments, all fabricated on a SiO$_2$ substrate. The operational principle of our device closely mirrors biological neural signal processing (Fig. 1a). Similar to how visual stimuli trigger neural responses through the human eye, optical input signals in our device propagate through defined pathways to generate controlled outputs. The input light wave enters through the waveguide port and sequentially couples to the first and second ring resonators through precisely engineered coupling regions (Fig. 1b,c). The optical input signal serves as the probe, while the controller corresponds to the pump, consistent with the pump-probe scheme described in prior studies [2,57].

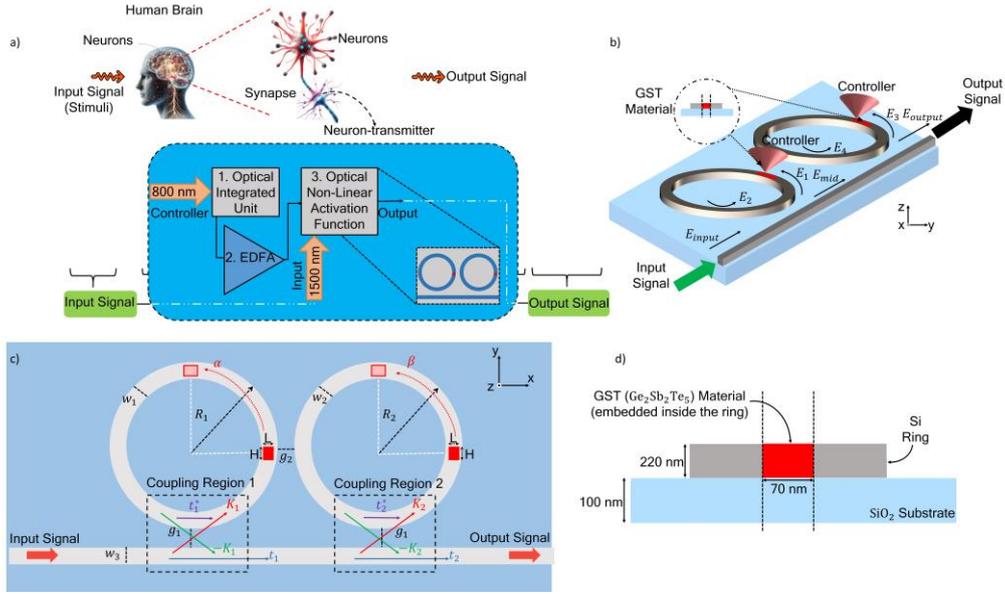

Fig. 1. a) Conceptual comparison between biological neural processing (top) and our photonic implementation (bottom), highlighting the analogous signal pathways and functional units. The optical integrated unit processes input signals through nonlinear activation functions implemented via ring resonators, mirroring biological neural signal processing, b) Three-dimensional schematic of the proposed double micro-ring resonator structure, showing the integrated GST segments and thermal controllers for precise phase-change modulation. The input signal (a.k.a probe) propagates through the waveguide, while the controller (a.k.a pump) drives the phase-change process. Directional light propagation ($E_{input}$ to $E_{output}$) illustrates the device's operational principle, c) Detailed top-view layout showing critical design parameters: ring widths ($W_1$, $W_2$), waveguide width ($W_3$), ring radii ($R_1$, $R_2$), coupling gaps ($g_1$, $g_2$), and angular positions ($\alpha$, $\beta$) of GST segments. The coupling regions ($t_1$, $t_2$, $k_1$, $k_2$) control light interaction between components, d) Cross-sectional view detailing the vertical structure: 220 nm-thick Si ring with embedded 70 nm GST material on SiO2 substrate, demonstrating the material integration strategy for optimal thermal and optical performance

The device's key structural parameters have been optimized for efficient operation. The ring widths $W_1$, $W_2$, and waveguide width $W_3$ are precisely controlled at 410 nm, 400 nm, and 410 nm, respectively. Both rings maintain an outer radius ($R_1$, $R_2$) of 5000 nm, with coupling gaps $g_1$ (100 nm) between the rings and waveguide and $g_2$ (400 nm) between the rings themselves. The embedded GST segments measure L = 70 nm in length and H = 500 nm in width, while the entire structure maintains a uniform thickness of 220 nm. A distinctive feature of our design is the angular positioning capability of the GST segments, denoted by angles $\alpha$ and $\beta$ in the first and second rings, respectively (Fig. 1c). This configuration enables precise control over the device's optical response through positional adjustments of the phase-change material. The light coupling behavior between components is governed by coupling coefficients $t_1$, $t_2$ (straight transmission) and $k_1$, $k_2$ (point coupling) at each coupling region. Detailed equations for resonance conditions and other key parameters, including Free Spectral Range (FSR), Full Width at Half Maximum (FWHM), and Quality Factor (Q), are provided in the Supplementary Material Section 1.

Using two distinct resonant wavelengths in our dual-ring resonator design is crucial for implementing reliable NLAFs. The first resonant wavelength is primarily tuned for efficient light coupling and high transmission, while the second wavelength ensures selective suppression of undesired optical modes. This dual-resonance configuration not only enhances the nonlinear control of optical signals by enabling precise modulation of transmission characteristics but also achieves high contrast ratios and robust switching. The two resonant wavelengths also provide flexibility for the device to operate across a broader range of input conditions, as each resonance can independently handle specific aspects of the input signal. Additionally, this dual-resonance structure supports multi-level optical processing, which is essential for emulating the weighted and dynamic signal processing observed in biological neurons. By enabling dual-switching functionality, the design facilitates complex, neuron-like behaviors and highlight significant potential for advancing the capabilities of neuromorphic photonic systems.

### 2.2 Angular Change of GST

Our device's tunability was investigated through a systematic analysis of sixteen distinct GST material configurations, illustrated in Fig. 2a-p. These configurations explore the impact of varying angular positions of GST within both ring resonators on the device's optical characteristics. The study encompasses four primary configuration sets, each demonstrating different combinations of angular shifts in the first and second rings.

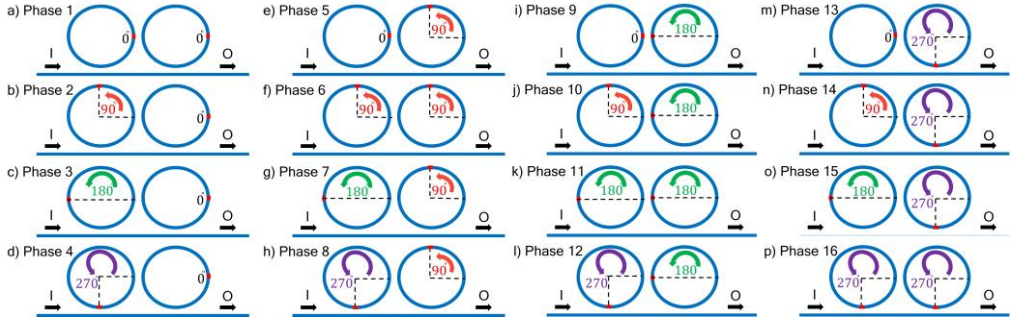

Fig. 2. a-p) Cross-sectional representation illustrating the phase transition dynamics of a 220 nm-thick silicon ring structure embedded with 70 nm GST material on a $SiO_2$ substrate, which includes the locations of the input port (I) and the Output port (O). The phases depict the angular positions of optical and thermal influences within the ring system, emphasizing the integration strategy for achieving optimized thermal and optical performance through precise phase control. a–d) GST material with no angular shift inside the second ring while exhibiting angular shifts inside the first ring, e–h) GST material with 90º angular shift inside the second ring, i–l) GST material with 180º angular shift inside the second ring, and m–p) GST material with 270º angular shift inside the second ring.

The first set (Fig. 2a-d) examines the effect of varying the GST position in the first ring from 90º to 270º while maintaining a fixed position in the second ring. In the second set (Fig. 2e-h), we introduce a 90º shift in the second ring while maintaining the same angular variation pattern in the first ring. The third set (Fig. 2i-l) increases the second ring's angular shift to 180º, and the fourth set (Fig. 2m-p) extends this to 270º, both while preserving the systematic variation in the first ring. Through comprehensive analysis of these configurations, based on criteria detailed in Supplementary Material Section 3 (Table S1), we identified the seventh phase configuration (Fig. 2g) as optimal, exhibiting superior performance with minimal transmission coefficients. This optimal configuration features a 180º GST material position in the first ring combined with a 90º position in the second ring. Following this discovery, we conducted detailed investigations of the second ring's width variations to further optimize the device's performance.

## 3. Numerical Calculations and Results

### 3.1. Angular Change Analyzes

We investigated the transmission characteristics of our device by calculating transmission coefficients for all sixteen phases at 25ºC, where the GST material is at an amorphous state, 100ºC where the GST material starts to change from amorphous to the crystalline state, 200ºC, where the GST is semi-crystalline, and 350ºC, which the GST material is at full crystalline state. In this configuration, presented in Fig. 2, all GST segments are assumed to be in the same state, either fully amorphous or fully crystalline, to evaluate the impact of angular positioning on the device's transmission characteristics under static conditions. The phase-change properties of GST are leveraged for optical switching by enabling reversible transitions between amorphous and crystalline states in subsequent dynamic operations. These transitions significantly modify the material's optical constants, such as refractive index and extinction coefficient, which directly influence the transmission characteristics of the device. For instance, the amorphous state provides higher transmission, while the crystalline state introduces stronger attenuation due to increased absorption. The combination of GST's phase-change properties and the angular positioning effects observed in Fig. 3a-p reveals how the transmission coefficient varies with GST material positioning inside the ring resonators at 25ºC, where the GST material is at the amorphous state. Each configuration exhibits two distinct resonant wavelengths, corresponding to the first and second rings, which is a key feature for implementing reliable nonlinear activation functions in neuromorphic computing.

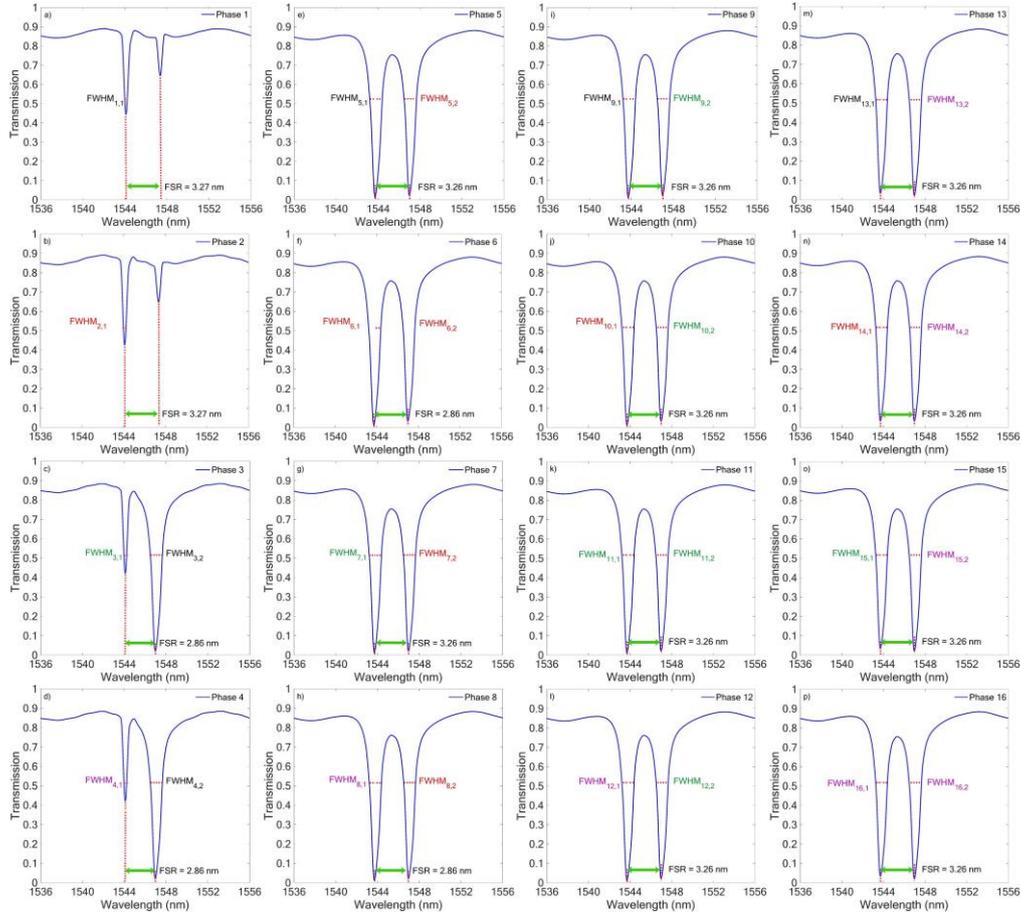

Fig. 3. Transmission spectra for sixteen different phase configurations of GST material embedded within the dual-ring resonator system as shown in Fig. 1 at 25ºC (Amorphous State). The spectra highlight the effects of varying angular shifts of the GST material in the first and second rings on the resonance characteristics. a–d) GST material in the second ring has no angular shift, while the first ring has angular shifts of 0º, 90º, 180º, and 270º, respectively. e–h) GST material in the second ring has a 90º angular shift, with corresponding angular shifts in the first ring of 0º, 90º, 180º, and 270º. i–l) GST material in the second ring has a 180º angular shift, with angular shifts in the first ring of 0º, 90º, 180º, and 270º. m–p) GST material in the second ring has a 270º angular shift, with angular shifts in the first ring of 0º, 90º, 180º, and 270º. The full-width half-maximum (FWHM) and free spectral range (FSR) values are annotated for each phase, illustrating the tunability and interplay between optical resonances within the system.

Phase 1 (Fig. 1a) shows resonant wavelengths at 1544.1 nm and 1547.37 nm, with transmission coefficients of 0.441 and 0.643, respectively. With a 180º angular shift in the first ring at Phase 3, the second resonant wavelength shifts to 1546.96 nm, yielding transmission coefficients of 0.0196 and 0.418. This dramatic change in transmission demonstrates the strong influence of GST positioning on light propagation within the rings. In Phase 4, where the GST material has a 270º shift, we observe transmission coefficients of 0.418 and 0.021, demonstrating how angular positioning can be used to selectively suppress specific resonant modes. In the 90º Second Ring Configuration (Phases 5-8, Fig. 3e-h), the resonant wavelengths stabilize at 1543.69 nm and 1546.96 nm, indicating a more consistent optical path length. Phase 6, where the GST material has a 90º shift in both rings, achieves low transmission coefficients of 0.011 and 0.0298, showcasing enhanced optical switching capability. Phase 7, where the GST material has a 180º shift in the first ring and a 90º shift in the second ring, achieves near-zero and 0.0184 transmission coefficients, providing the sharp contrast needed for effective neuromorphic switching operations. In the advanced configurations, Phases 9 to 12, as shown in Fig. 3i-p, the GST material in the second ring has a 180º shift and in the first ring changes from 0º to 270º. This behavior highlight the potential for reliable wavelength-selective switching. A 270º shift in the second ring from Phases 13 to 16 results in consistently near-zero transmission coefficients, indicating complete optical isolation. This feature is desirable for preventing crosstalk in integrated photonic circuits. A summary of all parameters obtained for sixteen different phases that represent the angular change of the GST material from 0º angle to 270º angle at 25 ºC, can be found in Table S1 from Supplementary Material Section S3. Also, Figure S1 in the Supplementary Material Section S3, illustrates how the transmission coefficient varies with GST material positioning inside the ring resonators at 100 ºC, where the state of the GST material changes from amorphous to the crystalline state. Fig. 4 illustrates the transmission coefficient and FWHM, which obtained from Fig. 3, and Table S1, for sixteen different phases at 25 ºC.

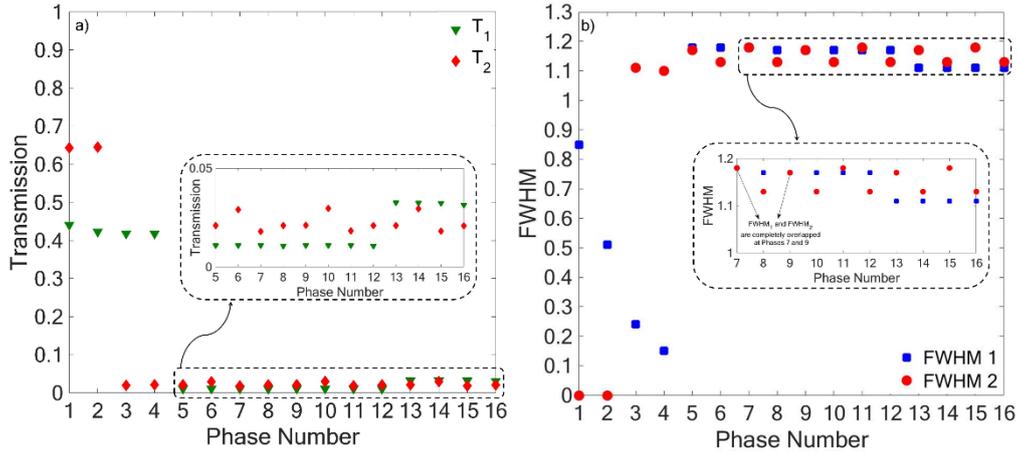

Fig. 4. Comparison of transmission coefficients and full-width half-maximum (FWHM) for the sixteen GST phase configurations in the dual-ring resonator system at 25 ºC. a) Transmission coefficients $T_1$ (green triangles) and $T_2$ (red diamonds) for the first and second resonance modes, respectively, across all phase configurations. The plot highlights significant variation in $T_1$ for early phases, while $T_2$ remains relatively low across most configurations. b) FWHM values for the first (blue squares) and second (red circles) resonance modes across all phases, illustrating distinct broadening behavior in specific configurations, with higher FWHM values observed for later phases.

The transmission characteristics in Fig. 4a demonstrate two distinct coefficients ($T_1$ and $T_2$) representing light propagation through the first and second ring resonators, respectively. The steady decline in transmission coefficients across phase configurations ($T_1$~0.4, $T_2$~0.65 in phases 1-4, dropping to near-zero in later phases) indicates increasingly effective optical mode confinement and GST-mediated absorption as the angular positions are optimized. The spectral characteristics, quantified by FWHM (Fig. 4b), expose a fundamental design trade-off inherent to ring resonator systems. Phases 1-4 achieve exceptionally narrow spectral widths (FWHM$_1$ ~0.2) due to strong resonant confinement, but their high transmission coefficients make them unsuitable for switching applications where high on-off contrast is essential. This behavior suggests that while these phases achieve excellent wavelength selectivity through constructive interference in the ring cavities, they fail to fully exploit GST's phase-change properties for transmission modulation. Notably, according to the provided data from Table S1, phases 7, 8, 11, and 12 strike an optimal balance between competing requirements. These configurations maintain consistent FWHM$_2$ values around 1.1 to 1.2, indicating stable, resonant behavior while simultaneously achieving the low transmission coefficients necessary for effective switching. This performance optimal point arises from the precise angular positioning of GST segments that maximizes light-matter interaction without compromising the resonant nature of the cavities. Temperature-dependent characterization at 100 ºC and 25 ºC, which are reported in Supplementary Material Section S3, Table S1, and Table S2, respectively, confirms phase seven as the ideal configuration. This phase maintains superior performance metrics across temperature variations, demonstrating robust switching behavior essential for practical neuromorphic computing applications. The resilience of phase seven to thermal variations suggests optimal alignment between the GST segments' optical properties and the resonant modes of the coupled ring system.

3.2 Tolerance Analyzes

To optimize device performance and understand operational constraints, we conducted detailed investigations of structural and thermal parameters. Our analysis focused on two key aspects which are dimensional variations in the ring resonators and temperature-dependent behavior of the GST material. We systematically investigated the impact of the second ring's width variations on device performance, focusing on phase seven, where GST materials in the first and second rings are positioned at 180º and 90º respectively (Fig. 2). The second ring was selected for analysis to avoid redundancy, as both rings are structurally identical. Therefore, any impact observed in the second ring's width is equally valid for the first ring. This choice simplifies the discussion while maintaining the generality of the results. Moreover, the analysis focuses on the second ring as a representative case since any changes in one ring would proportionally influence the coupled resonant modes of the dual-ring system. This symmetry ensures that the findings are comprehensive for the entire device. The width was analyzed to determine its tolerance range and consider how variations could affect device stability and reliability. This is crucial in practical applications, where fabrication inaccuracies may change the ring's dimensions and degrade performance.

The width of the second ring was varied from 350 nm to 450 nm while maintaining other structural parameters: first ring width $W_1$ (410 nm), waveguide width $W_3$ (410 nm), ring radii $R_1$ and $R_2$ (both 5000 nm), coupling gaps $g_1$ (100 nm) and $g_2$ (400 nm), and GST segment dimensions L (70 nm) and H (500 nm). The thickness of the ring and the rectangular waveguide is 220 nm. Detailed transmission characteristics for these variations are provided in Supplementary Material Section 4 in Figure S2 and Table S3. Also, Figure S3 illustrates the transmission coefficient and FWHM, which are obtained from Figure S2 and Table S3, for 10 different structures, where the width of the second ring changes from 350 nm to 450 nm.

### 3.3. Temperature Analyzes

The temperature-dependent behavior of our device reveals critical insights into its switching capabilities and operational stability. Fig. 5 confirms how temperature variation fundamentally alters the device's transmission characteristics through GST's phase-change properties. As temperature increases from 0°C to 350°C, the GST material undergoes significant changes in its optical properties, with its refractive index shifting from 3.25+0.09i to 5.6+0.85i54 [58].

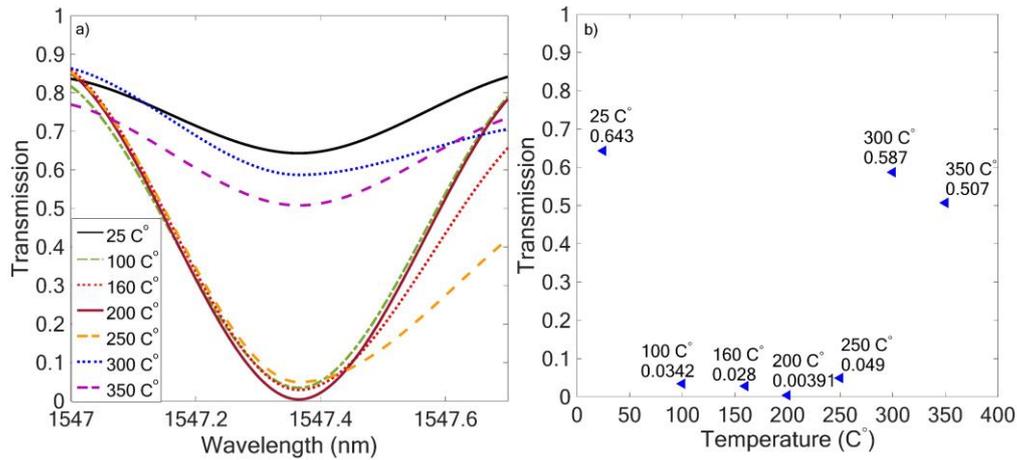

Fig. 5. Temperature-dependent transmission characteristics of GST material in the dual-ring resonator system for the first phase. a) Transmission spectra at wavelengths near 1547 nm for temperatures ranging from 25°C (Amorphous State) to 350°C (Crystalline State), showing significant changes in resonance depth and shape as the temperature increases. b) Transmission coefficient at 1547 nm as a function of temperature for the first phase configuration, highlighting a decrease in transmission with increasing temperature. The transition from amorphous to crystalline states of GST material is evident, with a notable reduction in transmission beyond 100°C.

Fig. 5a illustrates the evolution of transmission spectra across multiple temperatures (25°C to 350°C), revealing distinct behavioral regimes for the first phase. At room temperature (25°C), the device exhibits a broad transmission peak centered around 1547.1 nm with a maximum coefficient of approximately 0.85. As the temperature increases, we observe a systematic transformation in the transmission spectrum. The primary transmission dip deepens and shifts slightly toward shorter wavelengths. The spectral profile becomes more pronounced and asymmetric. Additional secondary features appear at longer wavelengths. The temperature-dependent transmission characteristics, summarized in Fig. 5b, demonstrate exceptional switching contrast. The transmission coefficient undergoes significant modulation. It starts at 0.643 at 25°C in the amorphous state, drops to 0.00391 at 200°C, which represents the optimal switching point, partially recovers to 0.587 at 300°C, and stabilizes at 0.507 by 350°C. This behavior suggests a complex interplay between the phase transition of GST and the resonant modes of the ring structure. The sharp transition corresponds to the crystallization temperature of GST, where the material undergoes significant changes in its optical properties. The subsequent partial recovery in transmission at higher temperatures indicates a refinement of the crystalline structure, providing additional control over the optical response of the device.

Fig. 6 reveals the temperature-dependent switching characteristics of phase seven, demonstrating enhanced performance compared to the baseline configuration. The spectral response (Fig. 6a) shows distinct temperature-dependent transmission features across the 1546-1553 nm wavelength range. At room temperature (25°C), the device exhibits low transmission (0.0194) at 1547 nm, indicating strong optical confinement. As temperature increases to 200°C, we observe a dramatic shift in behavior with transmission peaking at 0.677, representing a remarkable 0.657 change in transmission coefficient. Fig. 6b) shows that the transmission coefficient at 25°C is 0.0194 and increases to 0.677 while the temperature is 200°C and then decreases to 0.654 at 250°C, and finally decreases to 0.591 at 350°C. The lower switching temperature (≈ 200°C) of phase seven not only improves energy efficiency but also enhances device longevity by operating well below GST's crystallization temperature. The precise control over the spectral position and transmission amplitude demonstrates the potential for multi-level switching operations in neuromorphic computing applications.

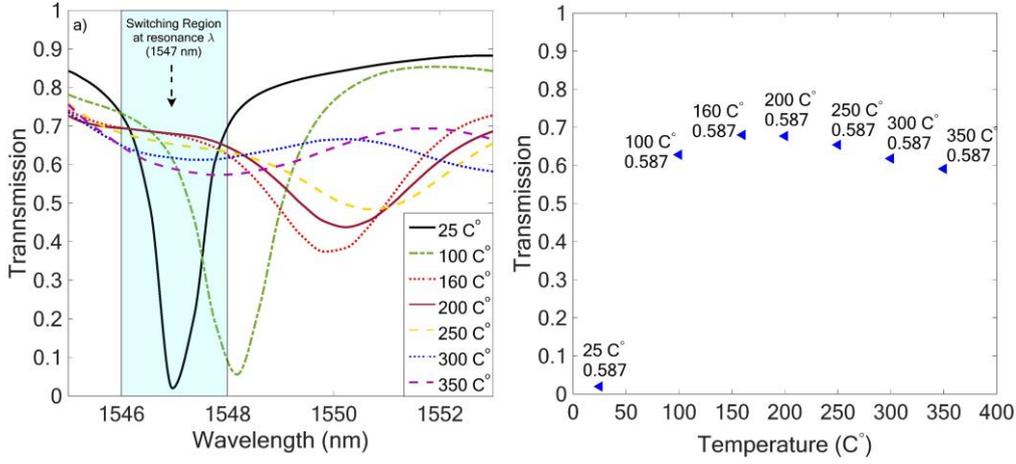

Fig. 6. Temperature-dependent transmission characteristics of GST material for the seventh phase configuration in the dual-ring resonator system. a) Transmission spectra over a wavelength range of 1546–1553 nm for temperatures between 25ºC (Amorphous State) and 350ºC (Crystalline State), showing significant spectral shifts and variations in transmission intensity as the temperature increases. b) Transmission coefficient at 1547 nm as a function of temperature, illustrating an increase in transmission with rising temperature, attributed to the gradual phase transition of the GST material from amorphous to crystalline states. The plot highlights key inflection points corresponding to intermediate and fully crystalline states of GST.

According to Fig. 5a and Fig. 6a, the maximum transmission variations are between 25 ºC and 200 ºC, so the transmission coefficient for the sixteen different phases is reported in Supplementary Material Figure S4 in Section 5.

The thermal behavior of our device, which is crucial for reliable switching operations, was characterized through detailed temporal and spatial analysis. Fig. 7 illustrates the cyclic temperature evolution during device operation, showing precisely controlled thermal excursions between room temperature, represented by the blue line, and the operational maximum of 350ºC. This temperature remains well below the melting point of GST, which is 600ºC and represented by the red line. The temporal temperature evolution follows a modified exponential relationship [57].

$$T(t) = T_0 + \Delta T_1 (1 - \exp(-\frac{t}{\tau_1})) - \Delta T_2 (1 - \exp(-\frac{t}{\tau_2})) \tag{1}$$

Where $T_0$ is the initial temperature, and here it is 25ºC, $\Delta T_1$ is the temperature changes while the temperature increases from the initial temperature to the peak value, called thermalization of the electrons and the lattice, which is 350ºC. Also, $\Delta T_2$ is the cooling temperature amplitude (350ºC), $\tau_1$, and $\tau_2$ are the heating and cooling rates, respectively. Equation 1 is adapted from a prior study investigating the time-domain separation of optical and structural transitions in GST. The exponential terms in the equation represent the distinct thermal processes, including rapid energy absorption by electrons (heating) and slower lattice relaxation (cooling). This functional form widely applies to GST-based systems, as it captures the underlying thermal mechanisms. In our work, the heating rate ($\tau_1$) and cooling rate ($\tau_2$) were extracted through numerical simulations incorporating the material properties of GST, such as thermal conductivity and heat capacity, along with the specific geometry of our structure. The heating rate for our simulation is 0.001 ns, and the cooling rate is 20% to 50%, and it is 0.1 ns. The fitted temperature curves align with the dynamics observed in the reference, validating the applicability of this model to our system. With rapid heating and controlled cooling, this asymmetric thermal response enables efficient switching while preventing thermal degradation.

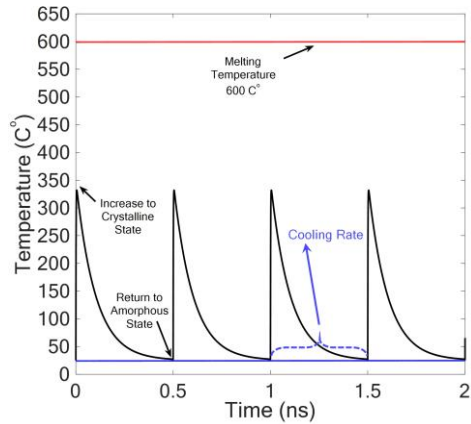

Fig. 7. Temporal temperature profile of GST material embedded within the microring resonators under periodic excitation from an optical source. The plot illustrates the rise and fall of the GST temperature over time, with peaks approaching the material's melting temperature of 600°C (indicated by the red dashed line) and the baseline temperature near ambient conditions (blue dashed line). These temperature cycles demonstrate controlled heating and cooling.

The device robust foundation for advancing this technology remarkable thermal cycling capability with a 0.5 ns cooling period, enabling switching rates up to 2 GHz. Each thermal cycle maintains precise control between the operational extremes, as evidenced by the consistent peak temperatures and baseline recovery shown in Fig. 7. The temperature-dependent transmission characteristics for all 16 GST phase configurations, including the significant changes in transmission coefficients between 25°C and 200°C, are detailed in Supplementary Material Figure S4 in Section 5.

The spatial heat distribution, as shown in Fig. 8, reveals distinct thermal profiles at operational extremes. At 350°C, illustrated in Fig. 8a, intense localized heating occurs within the GST segments, with controlled thermal spreading ensuring sharp switching transitions. At 25°C, illustrated in Fig. 8b, minimal residual heating is observed, confirming complete thermal recovery and system readiness for subsequent switching cycles. This thermal characterization confirms that the device is capable of reliable high-speed switching while maintaining thermal stability, which is a crucial requirement for neuromorphic computing applications.

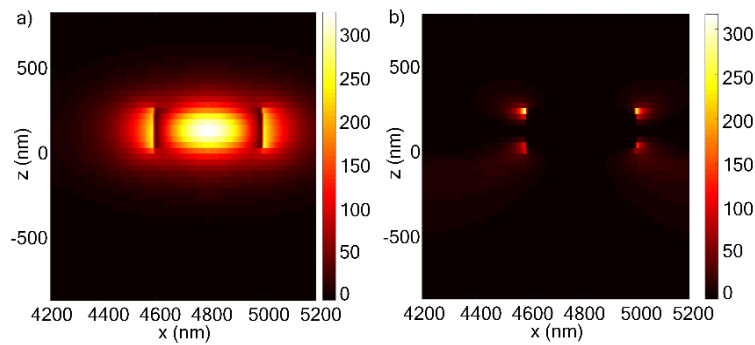

Fig. 8. Temperature distribution profiles of GST material embedded within the microring resonators. a) At 350°C (Crystalline State), the temperature distribution demonstrates a localized and intense heating region, indicative of the GST material approaching the phase transition threshold. The heat is concentrated at the center of the resonator, suggesting efficient energy confinement. b) At 25°C (Amorphous State), the temperature distribution shows minimal heating, with the GST material remaining in a stable, low-energy state. The contrast between the two profiles highlights the thermal response of the GST material under different operating conditions.

Fig. 9 reveals the fundamental optical properties of GST through its complex dielectric function ($\varepsilon = \varepsilon_1 + i\varepsilon_2$) in both amorphous and crystalline states, providing insight into the material's phase-dependent light-matter interactions.

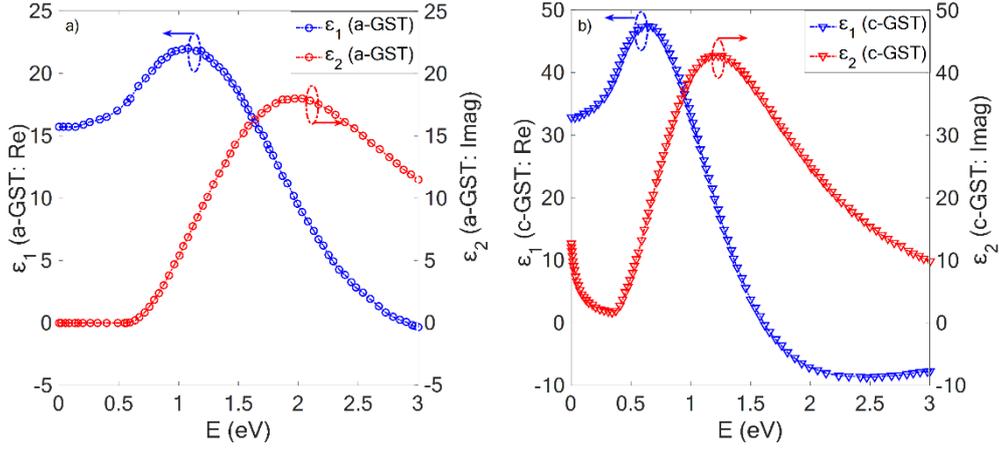

Fig. 9. Real ($\varepsilon_1$) and imaginary ($\varepsilon_2$) components of the dielectric function for GST material as a function of photon energy (E) in different states. a) In the amorphous state (a-GST), the dielectric function shows a distinct energy-dependent behavior, with $\varepsilon_1$ and $\varepsilon_2$ exhibiting moderate peaks, reflecting the less ordered atomic structure. b) In the crystalline state (c-GST), the dielectric function exhibits significantly higher peaks for both $\varepsilon_1$ and $\varepsilon_2$, indicating increased optical contrast and electronic transitions due to the more ordered atomic structure. The difference between a-GST and c-GST highlights the tunability of optical properties during phase transitions.

According to Fig. 9a, in the amorphous state, where the temperature is 25ºC, we observe distinct behavior across different energy regimes. In Fig. 9a, 0 to 0.5 eV is the low-energy region where the material exhibits minimal optical losses ($\varepsilon_2 \approx 0$) with a positive real permittivity ($\varepsilon_1 \approx 15$), indicating dielectric behavior. Also, 0.5 to 1.5 eV is the mid-energy region, where the imaginary component ($\varepsilon_2$) increases steadily while $\varepsilon_1$ reaches its maximum (~22), suggesting enhanced light-matter interaction. When the energy is greater than 1.5 eV, $\varepsilon_1$ decreases dramatically while $\varepsilon_2$ maintains significant values, indicating strong optical absorption. The crystalline state at 350ºC, as shown in Fig. 9b, demonstrates markedly different characteristics. The magnitude of both components is enhanced, with $\varepsilon_1$ peaking at approximately 48 and $\varepsilon_2$ reaching around 43. There are sharper transitions between energy regions and a notable transition to negative $\varepsilon_1$ values above 1.5 eV, indicating a metallic-like optical response. This phase-dependent evolution of dielectric properties enables our device's switching functionality by providing significantly different optical responses at our operating wavelength (1550 nm ≈ 0.8 eV). The transition from positive to negative $\varepsilon_1$ in the crystalline state is particularly significant when the energy exceeds 1.5 eV, as it fundamentally alters the material's interaction with incident light, enabling effective optical switching. The refractive index and dielectric function of GST material in both amorphous and crystalline states, critical for understanding the optical switching behavior, are analyzed in Supplementary Material Figures S6 and S7 in Section 7.

Table S4 in the Supplementary Material, Section 7, shows the real and imaginary parts of the refractive index of the GST material in its amorphous and crystalline states. Also, Table S5, in the Supplementary Material, Section 7, shows the dielectric function of the GST material in its amorphous and crystalline states at selected wavelengths between 255 nm and 1750 nm, and Table S6, summarizes the real and imaginary part of the permittivity of the GST material for both crystalline and amorphous states, while the energy changes from 0 eV to 3 eV.

*3.4. Power Analyzes*

Figure 10 presents the simulated output power spectra for two distinct GST phase configurations—Phase 1 (red curve) and Phase 7 (blue curve)—as a function of wavelength. The plotted range spans from approximately 1536 nm to 1555 nm, with output power varying between 0 mW and 3 mW. Both spectra exhibit multiple resonance dips and peaks, reflecting interference effects that arise due to the coupling dynamics between the waveguide and the dual-ring resonator system.

Key spectral differences between the two configurations highlight the impact of angular GST positioning on the resonance conditions. For instance, at around 1544.5 nm, Phase 1 exhibits a slightly higher output power (2.422 mW) compared to Phase 7 (2.24 mW), indicating stronger transmission through constructive interference. Conversely, at 1547.5 nm, Phase 1 reaches its peak power of 2.657 mW, outperforming Phase 7 in the same region, while at 1548.5 nm, Phase 7 exhibits higher output (2.296 mW) than Phase 1. These alternating patterns underscore the phase-dependent modulation of transmission characteristics.

The differences in peak locations and intensities can be attributed to variations in the effective refractive index caused by the GST phase state and angular placement. These changes alter the resonance conditions of the rings, modulating both the coupling efficiency and the constructive/destructive interference balance. Specifically, Phase 7—identified earlier as the optimal configuration—exhibits sharper, more defined resonance peaks due to better light confinement and stronger light-matter interaction at the GST boundaries. The slight reduction in output power is a trade-off for achieving higher spectral selectivity and switching contrast, which are critical for neuromorphic functionality.

Overall, these results reinforce the device's potential for optical switching and modulation. The phase sensitivity demonstrated here can be exploited for designing reconfigurable photonic circuits where output states are dynamically controlled by the GST phase, making this structure highly suitable for wavelength-selective neuromorphic operations.In the following, Fig. 10. illustrates the output power as a function of wavelength in phase one and phase seven.

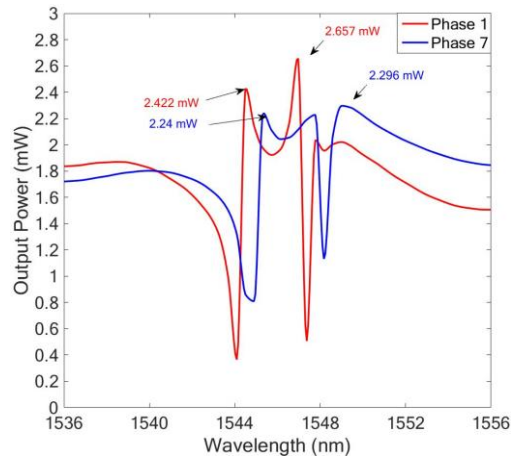

Fig. 10. Output power response of a proposed device as a function of wavelength for two different phase conditions, Phase 1 (red) and Phase 7 (blue). The spectral response exhibits multiple resonance dips and peaks due to interference effects within the coupled ring system. Phase variations induce shifts in the transmission characteristics, resulting in distinct power levels at specific wavelengths. Notable power values include 2.422 mW (Phase 1) and 2.24 mW (Phase 7) at approximately 1544.5 nm, as well as peak power differences at 1547.5 nm and 1548.5 nm. The results highlight the phase-dependent modulation of resonance properties, demonstrating the potential of the device for applications in optical switching, modulation, and neuromorphic photonics.

## 4. Discussion

Our work advances the field of neuromorphic computing by demonstrating how phase-change materials can be precisely engineered to emulate neural functionality at the hardware level. The developed GST-based double micro-ring resonator represents a fundamental building block for optical neural networks, offering several key advantages over traditional electronic implementations while more closely mimicking biological neural behavior.

The achievement of 0.5 ns switching speeds with our device architecture brings us closer to the temporal dynamics of biological neurons while operating at significantly lower power due to the optical domain implementation. More importantly, the ability to precisely control transmission characteristics through thermal and geometric parameters provides a pathway for implementing complex neural functions. The combination of ultra-narrow band transmission (FWHM of 0.47 nm) and high switching contrast mirrors the all-or-nothing firing behavior of biological neurons. At the same time, the continuous tunability enables the weighted signal processing essential for neural computation.

A particularly significant finding is how the angular positioning of GST segments influences device characteristics. The optimal configuration in phase seven balances switching contrast and spectral selectivity, analogous to how biological neurons maintain sensitivity while preventing spurious activation. This precise control over optical transmission, achieved through both material phase changes and geometric design, enables the implementation of nonlinear activation functions crucial for neural processing. The electric field distribution for Phase seven, illustrating the field confinement at resonant wavelengths, is provided in Supplementary Material Figure S5 (Section 6).

The temperature-dependent operation of our device (25ºC to 350ºC) demonstrates robust functionality well below GST's melting point, suggesting practical feasibility for large-scale integration. Achieving stable switching at relatively low temperatures (≈ 200ºC) with rapid thermal cycling (2 GHz capability) addresses one of the key challenges in neuromorphic hardware, which is maintaining reliability while operating at speeds relevant to real-world applications.

Looking beyond individual device performance, our work has broader implications for neuromorphic computing. The demonstrated integration of phase-change materials in photonic circuits provides a scalable approach for building dense neural networks. The double micro-ring architecture with precisely controlled GST angular positioning offers unprecedented control over switching characteristics while maintaining ultra-narrow band transmission (0.47 nm FWHM), enabling more precise neural activation functions than previous single-ring designs.

The achievement of multi-level switching capabilities through precise thermal control enables the implementation of synaptic plasticity. Our device achieves robust switching at lower operating temperatures (≈ 100ºC) than conventional GST implementations (≈ 200ºC) while maintaining high contrast ratios, a critical advancement for practical neural network integration.

The combination of high switching speeds and spectral selectivity opens possibilities for wavelength-division multiplexing in neural networks. Our design's distinctive combination of angular GST positioning and thermal control enables independent optimization of spectral selectivity and switching contrast, offering new degrees of freedom for neural response tuning.

The low thermal budget and efficient switching behavior suggest the potential for energy-efficient, large-scale neural systems. The demonstrated 0.5 ns cooling time with our optimized thermal management approach enables switching rates up to 2 GHz while maintaining stability well below GST's melting point, addressing key speed-reliability trade-offs in neuromorphic systems.

The double micro-ring resonator consumes slightly more area than single-ring designs; however, its enhanced spectral control, switching contrast, and nonlinear activation function justify the trade-off. Our design's compact radii (5 μm) significantly mitigate the footprint impact. Furthermore, the dual-ring architecture enables independent optimization of key performance metrics, making it a more versatile and effective solution for neuromorphic photonic applications.

These capabilities, particularly the combination of precise spectral control and efficient thermal switching, establish a new approach to implementing optical neural functions that could overcome scalability and energy efficiency limitations. While existing photonic neuromorphic systems often sacrifice either switching speed or spectral selectivity, our design demonstrates that both can be simultaneously optimized through careful consideration of material positioning and thermal management.

The success of our GST-based double micro-ring resonator architecture opens exciting pathways for further development of neuromorphic photonic systems. This thermal control strategy and angular switching precision lay a strong foundation for future device integration. Building upon our validated switching approach, future work can extend these principles to implement additional neural functionalities, leveraging the demonstrated thermal stability and spectral precision. The proven design principles can be scaled to create larger networks that take full advantage of the achieved ultra-narrow band transmission characteristics. Our established thermal management techniques enable opportunities for coordinated operation of multiple resonator elements. At the same time, our high-speed optical switching capabilities can be integrated with existing control systems to maximize practical impact.

These directions leverage the fundamental advances demonstrated in our work, particularly the achievement of precise switching control through optimized GST positioning and thermal management, to advance the field of neuromorphic photonic computing. Our results establish a clear path forward for realizing practical, scalable optical neural networks, bringing us closer to the goal of efficient, brain-inspired computing architectures. This work represents a significant step toward realizing practical neuromorphic systems that can approach the efficiency and capability of biological neural networks. Demonstrating precise control over neural-like behavior in a photonic platform establishes a promising direction for future brain-inspired computing architectures that could fundamentally transform how we process information.

## 5. Conclusion

We have introduced and validated a novel neuromorphic photonic architecture based on dual micro-ring resonators embedded with angularly configured GST segments. This approach enables precise, energy-efficient optical switching with tunable nonlinear activation characteristics, essential for next-generation optical neural networks.

Our optimized design achieves ultra-narrowband transmission (FWHM = 0.47 nm), high contrast modulation (0 to 0.85), and fast thermal switching at low operational temperatures (~100 °C). These features are enabled by independently tuning the spectral and switching properties of each resonator, a key advantage over previous single-ring systems. Through detailed simulation and thermal modeling, we demonstrated robust performance across a wide temperature range, fast switching cycles (0.5 ns), and compatibility with dense photonic integration. These attributes make the device well-suited for practical neuromorphic computing platforms. In summary, our results establish a scalable and efficient strategy for implementing nonlinear optical functions using phase-change materials. By leveraging geometric and thermal control in dual-ring systems, this work paves the way for more adaptable and biologically inspired photonic computing architectures.

## 6. Methods

### 6.1 Material Properties

The GST material can be used in both of the crystalline and amorphous states. The refractive index of the GST material can be changed from 3.25+0.09i to 5.6+0.85i. The refractive index of the GST material was taken from [59] and are available online in the Supplementary Material. Also, the refractive index of the Si and $SiO_2$ are taken from [60,61].

### 6.2 Fabrication Process

$SiO_2$ is a wafer layer, and 220 nm-thick layer of Si rectangular waveguide and ring resonators are placed on substrate layer with beam evaporation and lift-up process. Finally, ring resonators etched with exposing UV light and inserting nano-mask to embed GST layers inside the ring resonators. The length and width of the etching area is 70 nm and 500 nm, respectively. 220 nm-thick GST materials can be embedded with thermal evaporation inside the ring resonators.

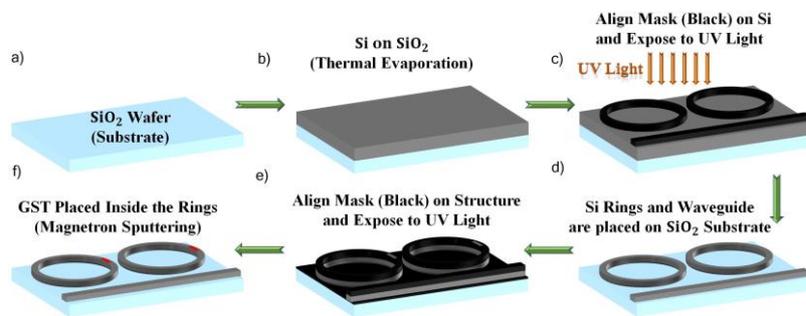

Fig. 11. Schematic representation of the fabrication process for a proposed device. a) $SiO_2$ wafer placed as a substrate layer. b) silicon layer is deposited on the substrate using thermal evaporation. c) A photolithography process is performed by aligning a mask onto the silicon layer and exposing it to UV light to define the structures. d) The silicon ring resonators and waveguide are patterned onto the $SiO_2$ substrate. e) A second photolithography step is conducted by aligning another mask on the structure and exposing it to UV light for further pattern refinement. f) The phase-change material (GST) is deposited inside the rings using magnetron sputtering, completing the fabrication process.

Fig. 11. illustrates the fabrication process of a proposed structure. The process begins from a $SiO_2$ wafer. In the next step, a thin layer of Si is deposited over the $SiO_2$ substrate via thermal evaporation. In the third step, the process of photolithography is carried out, where a mask (black) is aligned over the silicon layer and UV exposed for patterning the photonic structures. This exposure to UV allows selective etching of the silicon layer, thereby forming silicon ring resonators and a waveguide on the $SiO_2$ substrate. A second photolithography step is then performed by aligning and exposing a second mask to UV light to further define the structure for accurate patterning of the optical structures. Finally, a thin layer of GST is deposited in the rings through magnetron sputtering, integrating the phase-change material into the photonic device.

### 6.3 Simulation

The finite element method is used to simulate the ring resonators with a structure of Si and GST material.

### 6.4 Data Availability

The data supporting the findings of this study are available within the article and its Supplementary Material. Additional data are available from the authors upon request. Raw data (Refractive index and dielectric function for the GST material) can be found online: GST Material Raw Data Refractive Index Epsilon.

### 6.5 Conflict of interest

The authors declare that they have no conflicts of interest related to this research. The study was conducted in an objective and unbiased manner, and the results presented herein are based on rigorous analysis and interpretation. The authors have no financial or personal relationships with individuals or organizations that could potentially bias the findings or influence the conclusions of this study.

# Supplementary Information

# Dual Micro-Ring Resonators with Angular GST Modulation: Enabling Ultra-Fast Nonlinear Activation for Neuromorphic Photonics

Hossein Karimkhani[1], Yaser M. Banad[1], and Sarah Sharif[1,2*]

[1]School of Electrical and Computer Engineering, University of Oklahoma, Norman, OK, USA, 73019

[2]Center for Quantum Research and Technology, University of Oklahoma, Norman, OK, USA, 73019

[*]Corresponding authors: s.sh@ou.edu

## S1. Transmission Matrix Calculations in Micro Ring Resonators

The transmission matrix calculates the transmission between the rectangular waveguide and ring resonators and can be calculated by (S1) and (S2), according to Figure 1c [1,2].

$$\begin{bmatrix} E_{mid} \\ E_1 \end{bmatrix} = \begin{bmatrix} t_1 & -ik_1 \\ -ik_1 & t_1^* \end{bmatrix} \begin{bmatrix} E_{input} \\ E_2 \end{bmatrix} \quad (S1)$$

$$\begin{bmatrix} E_{output} \\ E_3 \end{bmatrix} = \begin{bmatrix} t_2 & -ik_2 \\ -ik_2 & t_2^* \end{bmatrix} \begin{bmatrix} E_{mid} \\ E_4 \end{bmatrix} \quad (S2)$$

If we consider our system is operating without loss, the relation between coefficients can be calculated by (S3).

$$t_1^2 + k_1^2 = t_2^2 + k_2^2 = 1 \quad (S3)$$

For simplification $E_{input}$ is equal to 1, and $E_2$ can be calculated by the following equation:

$$E_2 = e^{\frac{j\omega 2\pi r_1}{c}} E_1 \quad (S4)$$

With exploring (S1) and (S4) we can rewrite a new equation for $E_2$:

$$E_2 = \frac{-k_1^*}{-t_1^* + e^{\frac{j\omega 2\pi r_1}{c}}} \quad (S5)$$

where, ω is the angular frequency, c is the phase velocity of the ring, and can be calculated by S6

$$c = \frac{c_0}{n_{eff}} \quad (S6)$$

Where $c_0$ is the speed of the light in vacuum, and $n_{eff}$ is the effective refractive index. Also, $2\pi r_1$ is the circumference of the ring, and $r_1$ is the radius of the first ring resonator. The angular frequency can be calculated by $kc_0$ where k is the wave number and can be calculated by S7.

$$k = \frac{2\pi}{\lambda} \quad (S7)$$

Additionally, it is possible to calculate $E_1$, $E_{mid}$, $E_3$, $E_4$, and $E_{output}$ from the following equations.

$$E_1 = \frac{-k_1^*}{1-t_1^* e^{\frac{j\omega 2\pi r_1}{c}}} \quad (S8)$$

$$E_{mid} = \frac{-1+t_1^* e^{\frac{j\omega 2\pi r_1}{c}}}{-t_1^* + e^{\frac{j\omega 2\pi r_1}{c}}} \quad (S9)$$

$$E_3 = \frac{-k_2^*}{1-t_2^* e^{\frac{j\omega 2\pi r_2}{c}}} \quad (S10)$$

$$E_4 = \frac{-k_2^*}{-t_2^* + e^{\frac{j\omega 2\pi r_2}{c}}} \quad (S11)$$

$$E_{output} = \frac{-1+t_2^* e^{\frac{j\omega 2\pi r_2}{c}}}{-t_2^* + e^{\frac{j\omega 2\pi r_2}{c}}} \quad (S12)$$

Also, transmission coefficients between output and input can be calculated by (S13) and (S14) [3].

$$T_{output} = \left|\frac{E_{output}}{E_{mid}}\right|^2 = \left|\frac{t_2 e^A - \sqrt{e^{-B}}}{t_2^* e^A - \sqrt{e^{-B}}}\right| \qquad (S13)$$

$$T_{mid} = \left|\frac{E_{mid}}{E_{input}}\right|^2 = \left|\frac{t_1 e^C - \sqrt{e^{-D}}}{t_1^* e^C - \sqrt{e^{-D}}}\right| \qquad (S14)$$

where, $A = -\alpha L_{rt_2}$ and $C = -\alpha L_{rt_1}$ are coupling power attenuation. Also, $B = \Phi_{rt_2}$ and $D = \Phi_{rt_1}$ are round-trip optical phase for the first and the second ring, respectively.

## S2. Ring Resonator Parameters Calculations

Moreover, the round trip length of the resonance and resonance wavelength can be calculated by Eq. (S15) and (S16) [3–5].

$$L = 2(\pi R + L_c) \tag{S15}$$

$$\lambda_r = L n_{eff} m^{-1} \tag{S16}$$

where R represents the radius of the ring, and $L_c$ is the length of the coupling. L and $n_{eff}$ show the effective length and effective refractive index respectively. The other important parameters in ring resonators are Full Width Half Maximum (FWHM), Free Spectral Range (FSR), Fitness, and Quality Factor, which can be calculated by Eq. (S17), (S18), (S19), and (S20), respectively [3,6]. The system with a small FSR denotes that the system can support multiple channels in the proposed wavelength range [7]. Also, it is important to design two unalike rings to achieve distinct resonant wavelengths [8].

$$FWHM = \frac{c}{n\pi L} arccos \frac{2at_1 t_2}{1 + a^2 t_1^2 t_2^2} \tag{S17}$$

$$FSR = Wavelength_{c2} - Wavelength_{c1} = \Delta\lambda = \frac{\lambda^2}{n_g L} \tag{S18}$$

Where $n_g$ is the group refractive index and can be calculated by $n_{eff} - \lambda \frac{\partial n_{eff}}{\partial \lambda}$ [3].

$$Fit = \frac{FSR}{FWHM} \tag{S19}$$

$$Q\ Factor = \frac{Wavelength_c}{FWHM} = \frac{\pi n_{eff} L_c}{\lambda} \frac{\sqrt{at_1 t_2}}{1 - at_1 t_2} \tag{S18}$$

Where c is the speed of the light in free space, n is the refractive index of the waveguide and ring, $Wavelength_{c2}$, and $Wavelength_{c1}$ are the wavelengths in which the transmission has a peak. The proposed ring resonator has a resonant condition and can be expressed as Eq. (S19) [1,3].

$$2\pi R n = m \lambda_m \tag{S19}$$

Where R is the radius of the ring, n is the refractive index of the waveguide, and $\lambda_m$ is the required wavelength.

## S3. Summary of GST Material Angular Change Transmission

Table S1 summarizes the results from Figure 3(a–p) at 25 °C. It shows that the minimum Free Spectral Range (FSR) for the proposed structures is 2.86 nm, corresponding to phases three and four, where the GST material has a 0° angular shift in the first ring and a 180° or 270° angular shift in the second ring, respectively. In phases five to sixteen, the transmission coefficients are near zero, and the FSR in all these phases is 3.27 nm. According to Table S1, the best performance is achieved in phase seven, which exhibits a full width at half maximum (FWHM) of 1.18 nm for both resonant wavelengths. The quality factor for this structure is 1,308.212 at resonant wavelengths of 1,543.69 nm and 1,546.96 nm.

Figure S1 illustrates how changing the position of the GST material within the ring resonators affects the transmission coefficient, and Table S2 summarizes these results at 100 °C. In this figure, the transmission coefficient at the waveguide output port is shown for the amorphous GST material. Two distinct resonant wavelengths are observed: the first is associated with the second ring and the second with the first ring.

- Figure S1(a) (Phase 1): The resonant wavelengths are 1,544.10 nm and 1,547.37 nm, with transmission coefficients of 0.159 and 0.0342, respectively.
- Figure S1(b) (Phase 2): The transmission coefficients are 0.211 and 0.423 at the same resonant wavelengths.
- Figure S1(c) (Phase 3): With a 180° angular shift in the first ring, the second resonant wavelength shifts to 1,548.18 nm. The transmission coefficients are 0.0341 for the second ring at this wavelength and 0.418 for the first ring at 1,544.10 nm.
- Figure S1(d) (Phase 4): With a 270° angular shift in the first ring, the first resonant wavelength shifts to 1,544.51 nm. The transmission coefficients are 0.753 for the first ring at this wavelength and 0.061 for the second ring at 1,548.18 nm.

Figures S1(e–h) (Phases 5–8) depict scenarios where the GST material in the second ring has a 90° angular shift:

- Figure S1(e) (Phase 5): The resonant wavelengths are 1,544.92 nm and 1,548.18 nm, with transmission coefficients of 0.100 and 0.00281, respectively.
- Figure S1(f) (Phase 6): With a 90° angular shift in both rings, the resonant wavelengths are 1,545.32 nm and 1,548.18 nm, and the transmission coefficients are 0.174 and 0.194, respectively.
- Figure S1(g) (Phase 7): With a 180° angular shift in the first ring, the resonant wavelengths remain at 1,544.92 nm and 1,548.18 nm, but the transmission coefficients are near zero, specifically 0.033 and 0.034.
- Figure S1(h) (Phase 8): With a 270° angular shift in the first ring, the resonant wavelengths are 1,545.32 nm and 1,548.18 nm, with transmission coefficients of 0.123 and 0.076.

Figures S1(i–l) (Phases 9–12) show cases where the GST material has a 180° angular shift in the second ring. The resonant wavelengths remain at 1,544.92 nm and 1,548.18 nm throughout these phases, and the angular shift of the GST material in the first ring influences the transmission coefficients.

Figures S1(m–p) (Phases 13–16) illustrate scenarios where the GST material has a 270° angular shift in the second ring. Again, the resonant wavelengths are 1,544.92 nm and 1,548.18 nm, and in all these phases, the transmission coefficients are near zero.

A comprehensive summary of all parameters obtained for the sixteen different phases—representing the angular change of the GST material from 0° to 270°—can be found in the Supplementary Information. Table S2 provides detailed results for these phases from Figure S1(a–p) at 100 °C.

Table S1. Summary of all parameters obtained from Figure 3(a–p) at 25 °C for 16 different phases

| Phase | $T_1$ | $T_2$ | $\lambda_1$ (nm) | $\lambda_2$ (nm) | $FWHM_1$ | $FWHM_2$ | FSR | $Fit_1$ | $Fit_2$ | $Q_1$ | $Q_2$ |
|---|---|---|---|---|---|---|---|---|---|---|---|
| 1 | 0.441 | 0.643 | 1544.1 | 1547.37 | 0.85 | - | 3.27 | 3.847059 | - | 1816.59 | - |
| 2 | 0.423 | 0.645 | 1544.1 | 1547.37 | 0.51 | - | 3.27 | 6.411765 | - | 3027.647 | - |
| 3 | 0.418 | 0.0196 | 1544.1 | 1546.96 | 0.24 | 1.11 | 2.86 | 11.91667 | 2.576577 | 6433.75 | 1391.081 |
| 4 | 0.418 | 0.021 | 1544.1 | 1546.96 | 0.15 | 1.1 | 2.86 | 19.06667 | 2.6 | 10294 | 1403.727 |
| 5 | 0.011 | 0.021 | 1543.69 | 1546.96 | 1.18 | 1.17 | 3.27 | 2.771186 | 2.794872 | 1308.212 | 1319.393 |
| 6 | 0.011 | 0.029 | 1543.69 | 1546.96 | 1.18 | 1.13 | 3.27 | 2.771186 | 2.893805 | 1308.212 | 1366.097 |
| 7 | 0.011 | 0.018 | 1543.69 | 1546.96 | 1.18 | 1.18 | 3.27 | 2.771186 | 2.771186 | 1308.212 | 1308.212 |
| 8 | 0.0106 | 0.0209 | 1543.69 | 1546.96 | 1.17 | 1.13 | 3.27 | 2.794872 | 2.893805 | 1319.393 | 1366.097 |
| 9 | 0.011 | 0.0211 | 1543.69 | 1546.96 | 1.17 | 1.17 | 3.27 | 2.794872 | 2.794872 | 1319.393 | 1319.393 |
| 10 | 0.011 | 0.0297 | 1543.69 | 1546.96 | 1.17 | 1.13 | 3.27 | 2.794872 | 2.893805 | 1319.393 | 1366.097 |
| 11 | 0.0109 | 0.0183 | 1543.69 | 1546.96 | 1.17 | 1.18 | 3.27 | 2.794872 | 2.771186 | 1319.393 | 1308.212 |
| 12 | 0.0106 | 0.0209 | 1543.69 | 1546.96 | 1.17 | 1.13 | 3.27 | 2.794872 | 2.893805 | 1319.393 | 1366.097 |
| 13 | 0.0326 | 0.0210 | 1543.69 | 1546.96 | 1.11 | 1.17 | 3.27 | 2.945946 | 2.794872 | 1390.712 | 1319.393 |
| 14 | 0.0324 | 0.0295 | 1543.69 | 1546.96 | 1.11 | 1.13 | 3.27 | 2.945946 | 2.893805 | 1390.712 | 1366.097 |
| 15 | 0.0323 | 0.0182 | 1543.69 | 1546.96 | 1.11 | 1.18 | 3.27 | 2.945946 | 2.771186 | 1390.712 | 1308.212 |
| 16 | 0.0314 | 0.0208 | 1543.69 | 1546.96 | 1.11 | 1.13 | 3.27 | 2.945946 | 2.893805 | 1390.712 | 1366.097 |

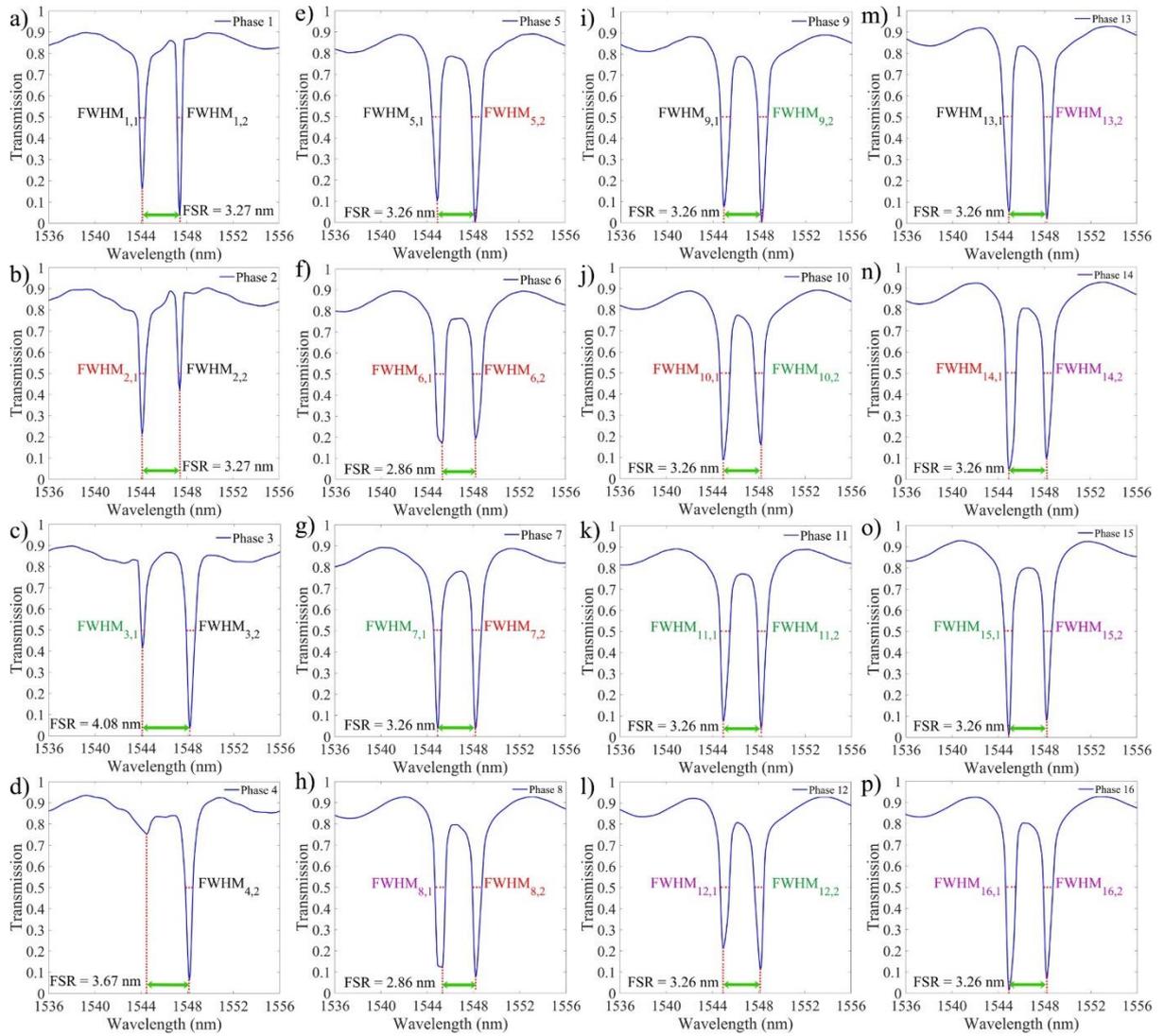

Figure S1. Transmission spectra at the waveguide output port for crystalline (100 °C) GST material in the dual-ring resonator system under sixteen different phase configurations. The plots show two distinct resonant wavelengths: one corresponding to the second ring and the other to the first ring. a-d) The GST material in the second ring has no angular shift, while the first ring exhibits angular shifts of 0°, 90°, 180°, and 270°, respectively. e-h) The GST material in the second ring has a 90° angular shift, with varying angular shifts in the first ring. i-l) The GST material in the second ring has a 180° angular shift, with the angular shift in the first ring varied across the four phases. m-p) The GST material in the second ring has a 270° angular shift, while the first ring's angular shift varies.

Each phase illustrates changes in resonant wavelengths, transmission coefficients, and full-width half-maximum (FWHM), highlighting the impact of GST material angular shifts on optical performance. Table S2, summarizes all of the results from Figure S1a-p) at 100 °C. Table S2 shows that the minimum FSR for the proposed structures is 2.86, which is related to phase six (GST material has 90° angular shift in both of the rings) and phase 8 (GST material has 90° and 270° angular shift in the first and the second ring, respectively). Also, the maximum amount of the FSR is 4.08, which is related to the third phase (GST material has 180° angular shift in the first ring). According to Table S2, in phases seven, nine, eleven, thirteen, fifteen, and sixteen, transmission coefficients are near zero. The FSR in all of these phases is 3.26. Table S2 demonstrates that the best results are related to phase seven, with 0.7 nm and 0.87 nm full-width half of maximum for the first and the second resonant wavelengths, respectively.

Also, the Quality factor for this structure is 2207.02 and 1779.51 at 1544.92 nm and 1548.18 nm resonant wavelengths.

Table S2. Summary of all parameters obtained from Figure S1(a–p) at 100 °C for 16 different phases

| Phase | $T_1$ | $T_2$ | $\lambda_1$ (nm) | $\lambda_2$ (nm) | $FWHM_1$ | $FWHM_2$ | FSR | $Fit_1$ | $Fit_2$ | $Q_1$ | $Q_2$ |
|---|---|---|---|---|---|---|---|---|---|---|---|
| 1 | 0.159 | 0.0342 | 1544.1 | 1547.37 | 0.85 | 0.47 | 3.27 | 3.84 | 6.95 | 1816.58 | 3292.27 |
| 2 | 0.211 | 0.423 | 1544.1 | 1547.37 | 0.51 | 0.14 | 3.27 | 6.41 | 23.35 | 3027.64 | 1105.64 |
| 3 | 0.418 | 0.0341 | 1544.1 | 1548.18 | 0.19 | 0.82 | 4.08 | 21.47 | 4.97 | 8126.842 | 1888.024 |
| 4 | 0.753 | 0.061 | 1544.51 | 1548.18 | - | 0.67 | 3.67 | - | 5.47 | - | 2310.71 |
| 5 | 0.1 | 0.00281 | 1544.92 | 1548.18 | 0.77 | 0.92 | 3.26 | 4.23 | 3.54 | 2006.38 | 1682.80 |
| 6 | 0.174 | 0.194 | 1545.32 | 1548.18 | 0.91 | 0.86 | 2.86 | 3.14 | 3.32 | 1698.15 | 1800.209 |
| 7 | 0.033 | 0.034 | 1544.92 | 1548.18 | 0.7 | 0.87 | 3.26 | 4.65 | 3.74 | 2207.02 | 1779.51 |
| 8 | 0.123 | 0.076 | 1544.32 | 1548.18 | 0.92 | 0.86 | 2.86 | 3.10 | 3.32 | 1679.69 | 1800.2 |
| 9 | 0.075 | 0.0023 | 1544.92 | 1548.18 | 0.89 | 0.91 | 3.26 | 3.66 | 3.58 | 1735.86 | 1701.29 |
| 10 | 0.0859 | 0.0159 | 1544.92 | 1548.18 | 0.92 | 0.83 | 3.26 | 3.54 | 3.92 | 1679.26 | 1865.27 |
| 11 | 0.0741 | 0.0394 | 1544.92 | 1548.18 | 0.89 | 0.84 | 3.26 | 3.66 | 3.88 | 1735.86 | 1843.07 |
| 12 | 0.214 | 0.109 | 1544.92 | 1548.18 | 0.79 | 0.85 | 3.26 | 4.12 | 3.83 | 1955.59 | 1821.38 |
| 13 | 0.05 | 0.0217 | 1544.92 | 1548.18 | 0.8 | 0.81 | 3.26 | 4.07 | 4.02 | 1931.15 | 1911.33 |
| 14 | 0.0479 | 0.099 | 1544.92 | 1548.18 | 0.92 | 0.89 | 3.26 | 3.54 | 3.66 | 1679.26 | 1739.52 |
| 15 | 0.0067 | 0.077 | 1544.92 | 1548.18 | 0.73 | 0.76 | 3.26 | 4.46 | 4.28 | 2116.32 | 2037.07 |
| 16 | 0.0148 | 0.0677 | 1544.92 | 1548.18 | 0.92 | 0.87 | 3.26 | 3.54 | 3.74 | 1679.26 | 1779.51 |

## S4. Transmission Coefficients for the Phase Seven with Different Width for the Second Ring Resonator

Figure S2a shows the transmission coefficients for Phase Seven when the width of the second ring is varied. At a width of 350 nm, the resonant wavelengths are 1545.73 nm and 1548.15 nm, with transmission coefficients of 0.0145 and 0.031, respectively. When the width is increased to 360 nm, the resonant wavelengths shift to 1548.18 nm and 1550.23 nm, with corresponding transmission coefficients of 0.027 and 0.0744. At 370 nm, the resonant wavelengths remain at 1548.18 nm and 1554.76 nm, with transmission coefficients of 0.028 and 0.0011, respectively.

Figure S2d demonstrates that at a width of 380 nm, the resonant wavelengths shift to 1537.61 nm and 1548.18 nm, with transmission coefficients of 0.239 and 0.026. Further increasing the width to 390 nm, 420 nm, and 430 nm results in resonant wavelengths of 1541.32 nm, 1548.15 nm, 1551.1 nm, and 1553.56 nm, with corresponding transmission coefficients of 0.044, 0.0106, 0.007, 0.0805, and 0.303.

Finally, for widths of 440 nm and 450 nm, the resonant wavelengths are 1548.15 nm and 1556.53 nm, and 1537.61 nm and 1548.18 nm, with transmission coefficients of 0.007 and 0.205, and 0.015 and 0.034, respectively. Table S3 summarizes all the results presented in Figure S2a-j, highlighting the impact of varying the second ring width on the transmission and resonance characteristics.

Figure S3a provides a detailed comparison of the first and second transmission coefficients for Phase Seven as the width of the second ring changes from 350 nm to 450 nm. Figure S3b illustrates the first and second FWHM values for the same phase and width range. It can be observed from Figure S3a that the transmission coefficients are lowest at a width of 370 nm, while Figure S3b shows that the best FWHM values are achieved at a width of 400 nm.

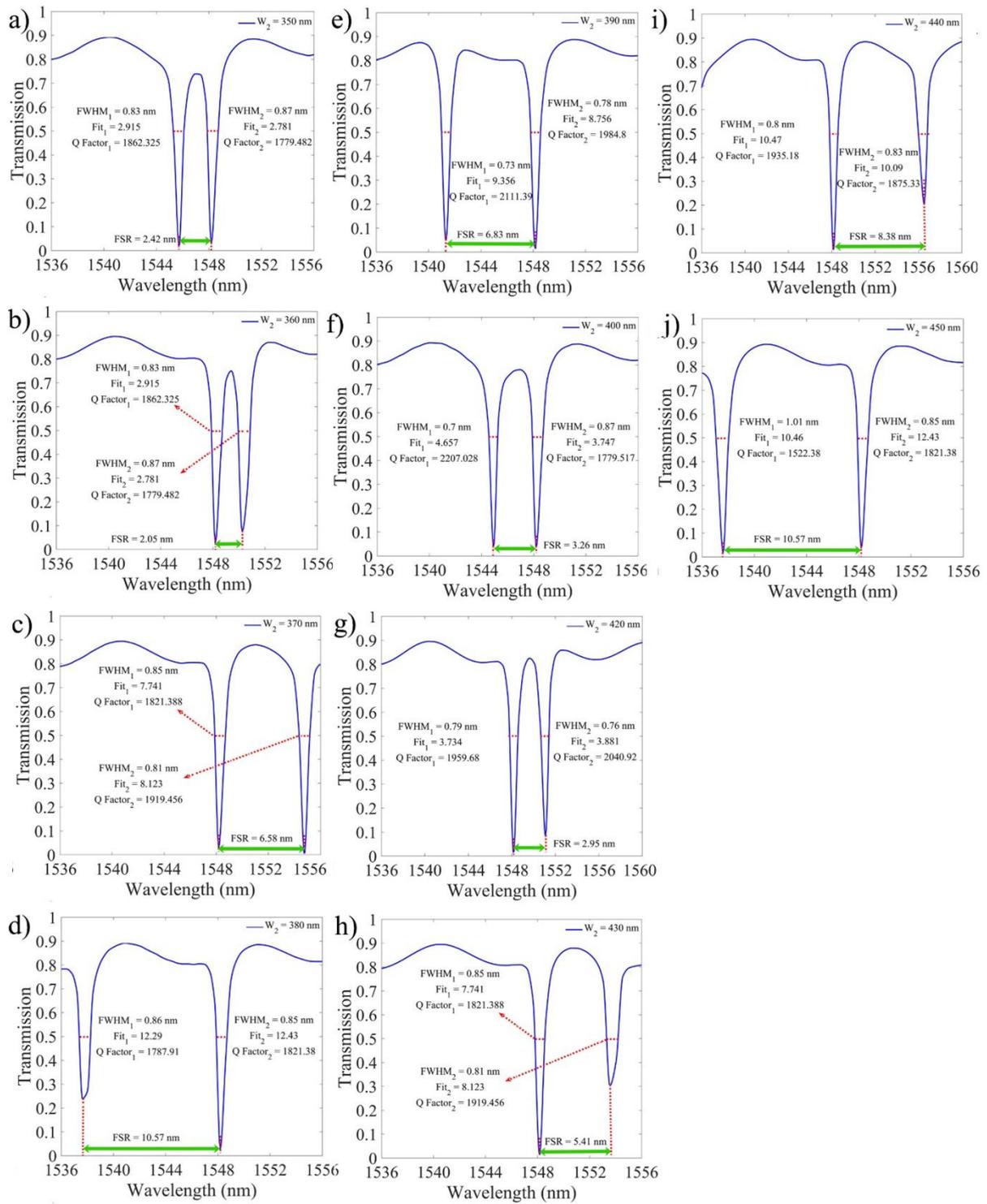

Figure S2. Transmission coefficients for the Phase seven, where the GST material inside the rings has 180° angular shift in the first ring resonator and 90° angular shift in the second ring resonator, while the width of the second ring changes from 350 nm to 380 nm. The width of the second ring is a) 350 nm, b) 360 nm, c) 370 nm, d) 380 nm, e) 390 nm, f) 400 nm, g) 420 nm, h) 430 nm, i) 440 nm, j) 450 nm

Table S3. Summary of all parameters obtained from Figure S2(a–j) while the width of the second ring at phase seven changes from 350 nm to 450 nm

| $W_2$ | $T_1$ | $T_2$ | $\lambda_1$ (nm) | $\lambda_2$ (nm) | $FWHM_1$ | $FWHM_2$ | FSR | $Fit_1$ | $Fit_2$ | $Q_1$ | $Q_2$ |
|---|---|---|---|---|---|---|---|---|---|---|---|
| 350 | 0.0145 | 0.031 | 1547.73 | 1548.15 | 0.83 | 0.87 | 2.42 | 2.915 | 2.781 | 1862.32 | 1779.48 |
| 360 | 0.027 | 0.0744 | 1548.18 | 1550.23 | 0.87 | 0.94 | 2.05 | 2.356 | 2.18 | 1779.517 | 1649.18 |
| 370 | 0.028 | 0.0011 | 1548.18 | 1554.76 | 0.85 | 0.81 | 6.58 | 7.741 | 8.123 | 1821.388 | 1919.456 |
| 380 | 0.239 | 0.026 | 1537.61 | 1548.18 | 0.86 | 0.85 | 10.57 | 12.29 | 12.43 | 1787.918 | 1821.38 |
| 390 | 0.044 | 0.0106 | 1541.32 | 1548.15 | 0.73 | 0.78 | 6.83 | 9.356 | 8.756 | 2111.397 | 1984.807 |
| 400 | 0.033 | 0.034 | 1544.92 | 1548.18 | 0.7 | 0.87 | 3.26 | 4.65 | 3.74 | 2207.02 | 1779.51 |
| 420 | 0.007 | 0.0805 | 1548.15 | 1551.1 | 0.79 | 0.76 | 2.95 | 3.734 | 3.881 | 1959.68 | 2040.92 |
| 430 | 0.01 | 0.303 | 1548.15 | 1553.56 | 0.79 | 0.84 | 5.41 | 6.84 | 6.44 | 1959.68 | 1849.47 |
| 440 | 0.007 | 0.205 | 1548.15 | 1556.53 | 0.007 | 0.205 | 8.38 | 10.47 | 10.09 | 1935.18 | 1875.33 |
| 450 | 0.015 | 0.034 | 1537.61 | 1548.18 | 1.01 | 0.85 | 10.57 | 10.46 | 12.43 | 1522.38 | 1821.38 |

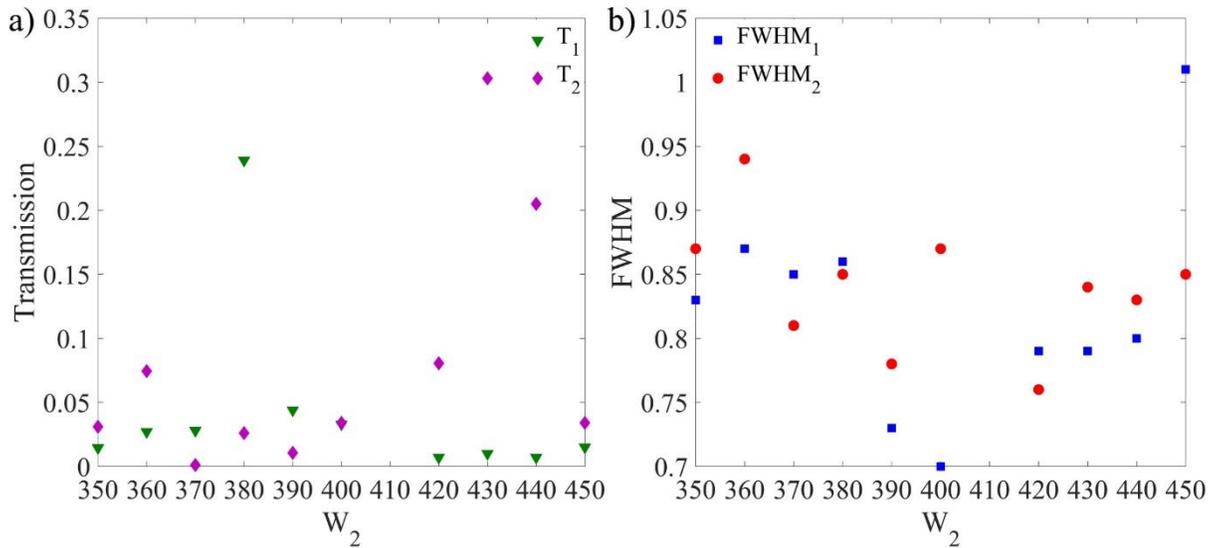

Figure S3. First and second a) transmission coefficient and b) FWHM for the seventh phase while the width of the second ring changes from 350 nm to 450 nm.

## S5. Transmission Coefficient for Sixteen Phases at Two Temperatures

Figure S4 illustrates the transmission coefficients for all sixteen phases at 25°C (blue curves) and 200°C (red curves), highlighting the temperature-dependent behaviour of the dual-ring resonator system.

- **Figures S4a-d**: These panels show the transmission coefficients for the first four phases, where the GST material in the second ring has no angular shift, and the first ring has angular shifts of 0°, 90°, 180°, and 270°. In Phase 1 (Figure S4a), the transmission coefficient undergoes a significant change when the temperature increases from 25°C to 200°C, demonstrating a notable response. However, in Figures S4b-d, as the angular shift in the first ring increases, the temperature-dependent changes in the transmission coefficients become broader and less distinct, with shifts of approximately 0.3, which may not be ideal for the proposed application.

- **Figures S4e-h**: These panels represent phases where the GST material in the second ring has a 90° angular shift. In Phase 5 (Figure S4e), the transmission coefficients at 25°C and 200°C remain nearly identical, showing minimal temperature dependence. For Phases 6-8 (Figures S4f-h), the GST material in the first ring has angular shifts of 90°, 180°, and 270°, respectively. While some changes in the transmission coefficients are observed at 200°C, the shifts are relatively minor, with a maximum of around 0.3, and the temperature response remains broad.

- **Figures S4i-l**: These panels illustrate phases with a 180° angular shift in the GST material in the second ring. Across these phases, the transmission coefficients show minimal variation between 25°C and 200°C, indicating weak temperature sensitivity.
- **Figures S4m-p**: These panels show phases where the GST material in the second ring has a 270° angular shift. Similar to the 180° angular shift, the transmission coefficients across these phases display negligible differences between the two temperatures, suggesting limited temperature responsiveness.

Overall, Figure S4 demonstrates that the first phase (Figure S4a), where the GST material has no angular shift in either ring, provides the best temperature response. In this phase, the transmission coefficient exhibits a substantial change from 0 to 0.7 as the temperature increases from 25°C to 200°C. This makes Phase 1 the most promising configuration for applications requiring significant temperature-dependent transmission changes.

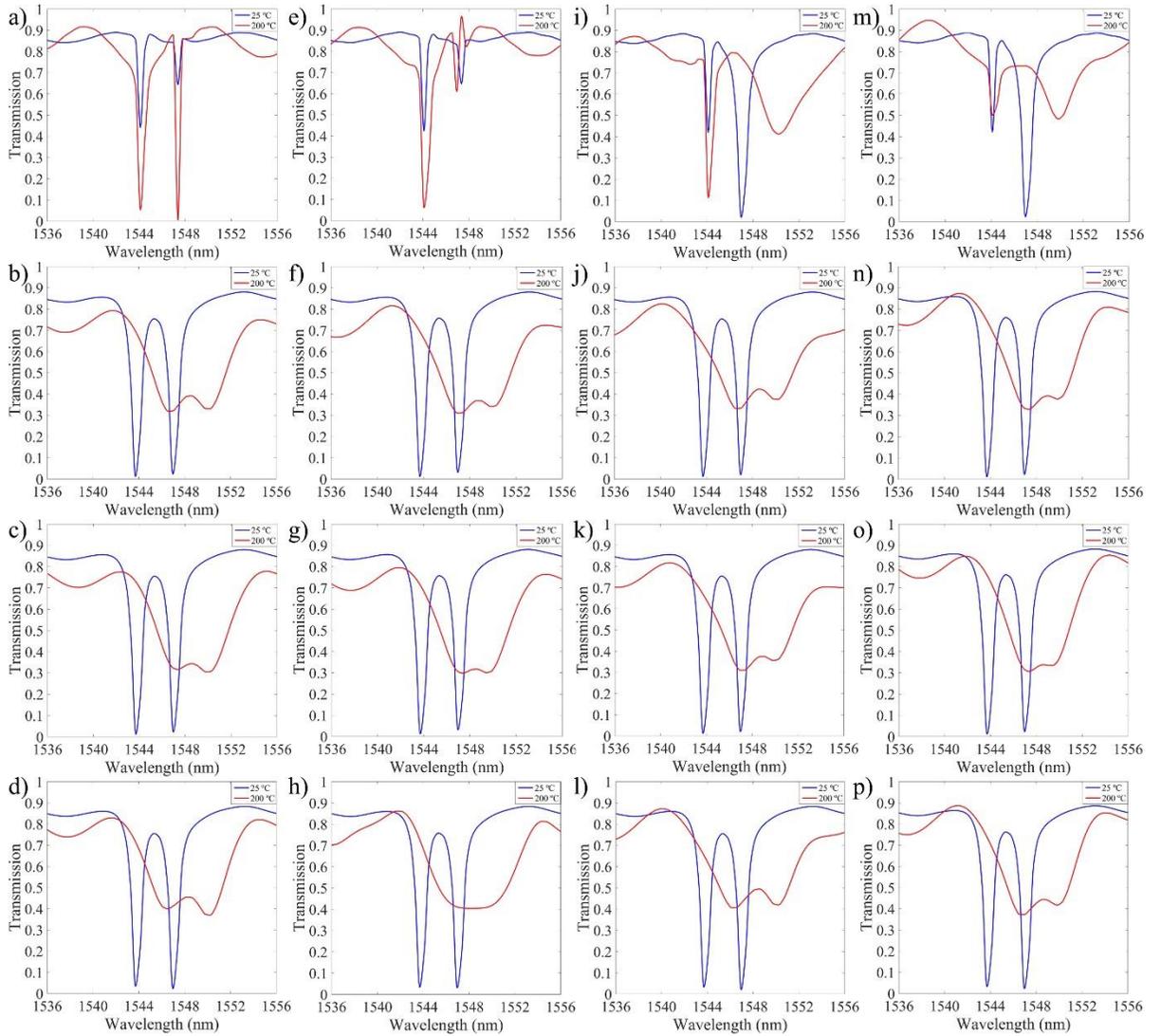

Figure S4: Transmission coefficient for all of the sixteen phases at 25° C and 250° C, where the GST materials inside the ring resonators have 0° to 270° angular shift.

Figure S5 illustrates the temperature-dependent transmission characteristics of GST material for the first and the seventh phase configuration for the first resonance wavelength. Figure S5a illustrates the transmission coefficient across multiple temperatures (25ºC to 350ºC) for the first phase at the first resonance wavelength, which is related to the other ring resonator's resonance. At room temperature (25ºC), the device exhibits a broad transmission peak centered around 1544.1 nm with a maximum coefficient of approximately 0.44. As temperature increases, we observe a systematic transformation of the transmission spectrum, which increases to 0.68 at 300ºC and drops to 0.15 at 250ºC in the first phase. Figure S5b illustrates the transmission coefficients at these temperatures. Also, Figure S5c shows the temperature-dependent transmission characteristics of GST material for the seventh phase at the other resonance wavelength. As temperature increases, we observe a systematic transformation of the transmission spectrum and increases to 0.73 at 200ºC in the seventh phase. Figure S5d demonstrates that with changing the temperature between 25ºC and 350ºC, the transmission coefficient alters between 0.43 and 0.73, showing a 0.33 variation of the transmission in switching. However, the first phase shows a 0.53 variation between 25ºC and 350ºC.

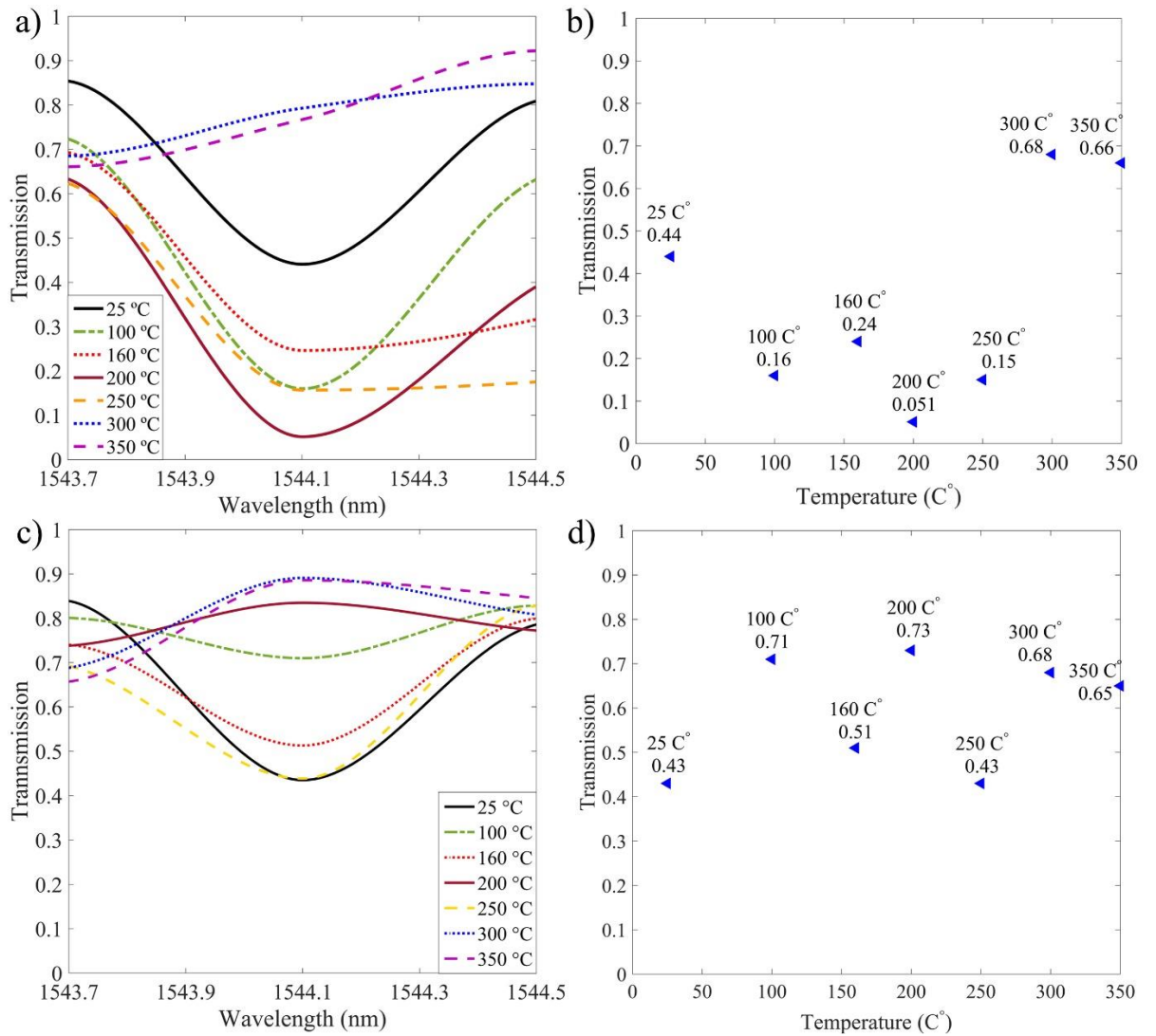

Figure S5: Temperature-dependent transmission characteristics of GST material in the dual-ring resonator system. a) Transmission spectra at wavelengths near 1544 nm for temperatures ranging from 25◦C to 350◦C, showing significant changes in resonance depth and shape as the temperature increases for the first phase. b) Transmission coefficient at 1544 nm as a function of temperature for the first phase configuration, highlighting an increase in transmission

with increasing temperature. The transition from amorphous to crystalline states of GST material is evident, with a notable promotion in transmission beyond 100◦C. c) Transmission spectra at wavelengths near 1544 nm for temperatures ranging from 25◦C to 350◦C, for the seventh phase, d) Transmission coefficient at 1544 nm as a function of temperature for the seventh phase configuration, which shows a small switching transmission changes.

## S6. Electrical Field Distribution for Phase Seven

Figure S6 illustrates the electrical field profile at various wavelengths. Figure S6a) illustrates the first and the second transmission coefficients for the seventh phases while the width of the second ring changes from 350 nm to 450 nm. Figure S3b shows the first and the second FWHM for the seventh phase for various widths of the second ring. Figure S6a) demonstrates that while the width of the second ring is 370 nm, the structure at the seventh phase has the lowest transmission coefficients; however, at 400 nm, it has the best FWHM.

Figure S6 illustrates the electric field distribution and corresponding transmission coefficients at two resonant wavelengths for Phase Seven. In **Figure S6a**, the transmission spectrum shows two distinct resonant wavelengths, $\lambda 1=1544.92$ nm, and $\lambda 1=1548.18$ nm, corresponding to the coupling regions in the dual-ring resonator. These wavelengths are highlighted with their respective transmission coefficients, demonstrating the resonator's filtering and coupling behavior. In **Figure S6b,** this panel shows the electric field profile at $\lambda 1=1544.92$ nm, where the electric field is strongly confined in **Coupling Region 1**. The energy is primarily concentrated within the first ring resonator, confirming efficient coupling between the input waveguide and the first ring. In **Figure S6c,** the electric field profile at $\lambda 1=1548.18$ nm highlights strong field confinement in **Coupling Region 2**, which is associated with the second ring resonator. The energy is transferred efficiently between the input waveguide and the second ring, showcasing distinct wavelength-selective behavior. The analysis reveals that the seventh phase provides clear field localization at the resonant wavelengths, enabling selective control of transmission properties. The resonant behavior and field confinement demonstrated in Figure S6 underscore the importance of phase and structural configuration for optimizing transmission performance in the dual-ring resonator system.

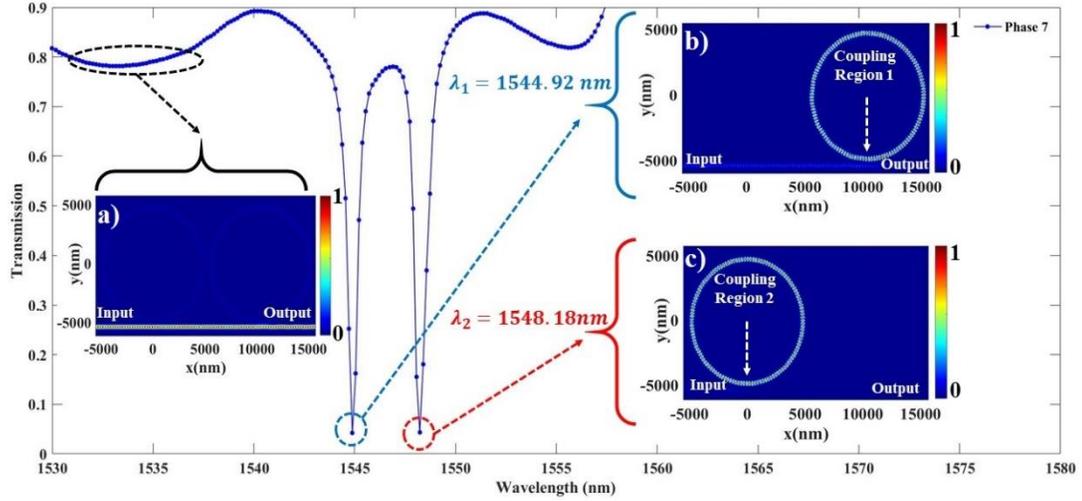

Figure S6. Electrical field distribution for phase seven at a) below 1544.92 nm wavelengths, which go entirely to the output through the port, at b) 1544.92 nm wavelengths, which are entirely confined inside the second ring resonator; and at c) 1548.18 nm wavelengths; which are confined in the first ring resonator.

## S7. Refractive index calculations for GST material

The refractive index of the GST material is derived from experimental data provided in Reference [9]. Using this data, the dielectric function of the GST material can be calculated with the following equation:

$$\varepsilon_{GST}(\omega) = (n_{GST}(\omega) + ik_{GST}(\omega))^2 \qquad (S20)$$

where, $n_{GST}(\omega)$ and $k_{GST}(\omega)$ represent the real and imaginary parts of the GST refractive index, respectively.

Figure S7 shows the variation in the real ($n$) and imaginary ($k$) components of the refractive index for both amorphous (25°C) and crystalline (200°C) states. The data illustrates a notable increase in both n and k as the temperature transitions from 25°C to 200°C across the wavelength range of 700 nm to 1800 nm. The effective permittivity $\varepsilon_{eff}(\omega)$ of GST material at different crystallization states can also be calculated using the Lorentz-Lorenz relation [10–12]:

$$\frac{\varepsilon_{eff}(\omega) - 1}{\varepsilon_{eff}(\omega) + 2} = p \frac{\varepsilon_{GST-C}(\omega) - 1}{\varepsilon_{GST-C}(\omega) + 2} + (1-p) \frac{\varepsilon_{GST-A}(\omega) - 1}{\varepsilon_{GST+A}(\omega) + 2} \qquad (S21)$$

where $\varepsilon_{GST-C}(\omega)$ is the permittivity of the GST material at crystalline state, and $\varepsilon_{GST-A}(\omega)$ is the permittivity of the GST material at the amorphous state. Also, p is the crystallization rate of the GST material, which ranges between 0 and 1. While the material is in the crystalline state, p is 1, and while the material is in the amorphous state, p is 0.

**Table S4** summarizes the real and imaginary components of the refractive index (n and k) for the GST material at various wavelengths for both crystalline and amorphous states. Similarly, **Table S5** provides a detailed breakdown of the dielectric function ($\varepsilon_1$ and $\varepsilon_2$) for the GST material at selected wavelengths.

In **Figure S8**, the real and imaginary parts of the dielectric function are plotted for both amorphous and crystalline states over a range of wavelengths. Beyond 800 nm, the difference between the two states becomes more pronounced, with the imaginary part showing significant increases in the crystalline state. This behaviour underscores the strong optical contrast achievable by transitioning between the two phases, which is critical for phase-change-based optical devices.

Table S6 summarizes the real ($\varepsilon_1$) and imaginary ($\varepsilon_2$) part of the permittivity of the GST material for both crystalline and amorphous states, while the energy changes from 0 eV to 3 eV.

Table S4. Refractive index of the GST material. Summary of the real ($n$) and imaginary ($k$) components of the refractive index for GST material in its amorphous (25°C) and crystalline (200°C) states at selected wavelengths (255 nm to 1750 nm).

| λ (nm) | 255 | 357 | 574 | 785 | 942 | 1110 | 1260 | 1420 | 1590 | 1750 |
|---|---|---|---|---|---|---|---|---|---|---|
| n-cGST | 0.744 | 1.53 | 3.23 | 4.78 | 5.74 | 6.59 | 7.09 | 7.23 | 6.91 | 6.11 |
| k-cGST | 2.49 | 3.29 | 4.2 | 4.32 | 4.08 | 3.58 | 3.02 | 2.4 | 1.86 | 1.5 |
| n-aGST | 1.69 | 2.59 | 3.85 | 4.52 | 4.76 | 4.78 | 4.71 | 4.61 | 4.51 | 4.47 |
| k-aGST | 2.59 | 2.49 | 2.09 | 1.57 | 1.13 | 0.754 | 0.436 | 0.177 | 0.0987 | 0.181 |

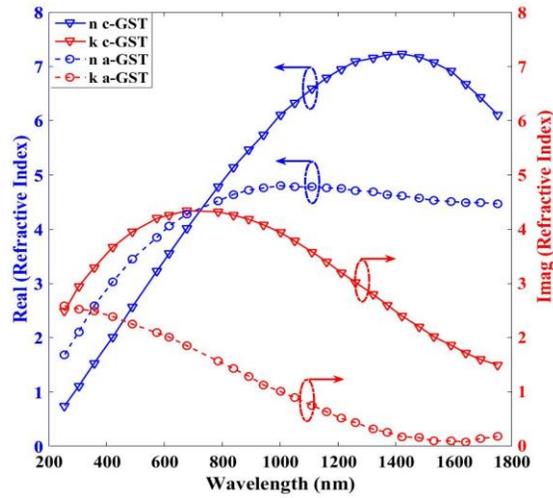

Figure S7. Refractive Index of GST Material. Real (n) and imaginary (k) parts of the refractive index for GST material in amorphous (25°C) and crystalline (200°C) states, plotted across the wavelength range of 700 nm to 1800 nm.

Table S5. Dielectric function of the GST material. Summary of the real ($\varepsilon_1$) and imaginary ($\varepsilon_2$) parts of the dielectric function for GST material in its amorphous (25°C) and crystalline (200°C) states at selected wavelengths (255 nm to 1750 nm).

| $\lambda$ (nm) | 255 | 357 | 574 | 785 | 942 | 1110 | 1260 | 1420 | 1590 | 1750 |
|---|---|---|---|---|---|---|---|---|---|---|
| $\varepsilon_1$-cGST | -5.64 | -8.48 | -7.20 | 4.18 | 16.3 | 30.61 | 41.15 | 46.51 | 44.29 | 35.08 |
| $\varepsilon_2$-cGST | 3.70 | 10.07 | 27.13 | 41.3 | 46.84 | 47.18 | 42.82 | 34.7 | 25.71 | 18.33 |
| $\varepsilon_1$-aGST | -6.15 | -3.85 | 6.065 | 20.38 | 31.67 | 42.86 | 50.08 | 52.24 | 47.74 | 37.3 |
| $\varepsilon_2$-aGST | 3.85 | 7.61 | 13.5 | 15.01 | 12.97 | 9.93 | 6.18 | 2.55 | 1.36 | 2.21 |

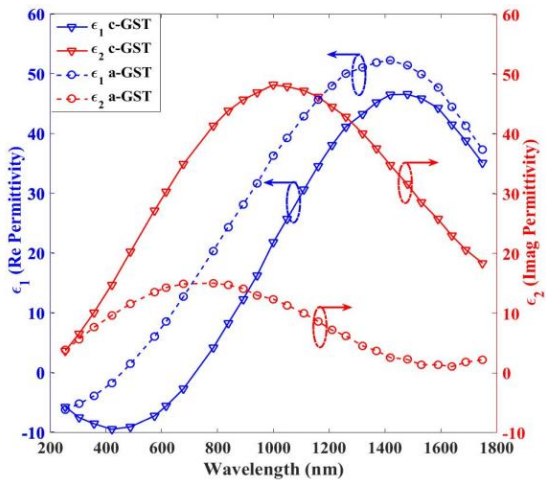

Figure S8. Dielectric Function of GST Material. Real ($\varepsilon_1$) and imaginary ($\varepsilon_2$) parts of the dielectric function for GST material in amorphous (25°C) and crystalline (200°C) states, plotted across the wavelength range of 700 nm to 1800 nm.

Table S6. Comprehensive summary of parameters for 16 different phases of GST, representing angular changes from 0° to 270° in the dual-ring resonator system.

| Energy (eV) | 0 | 0.5 | 0.75 | 1 | 1.5 | 2 | 2.26 | 2.5 | 2.75 | 3 |
|---|---|---|---|---|---|---|---|---|---|---|
| $\varepsilon_1 cGST$ | 32.8 | 45.2 | 45.7 | 33.1 | 3.78 | -7.08 | -8.45 | -8.65 | -8.35 | -7.67 |
| $\varepsilon_2 cGST$ | 12.7 | 6.83 | 25.4 | 39.2 | 38.2 | 24.7 | 19.3 | 15.4 | 12.3 | 9.87 |
| $\varepsilon_1 aGST$ | 15.7 | 17.5 | 19.9 | 21.9 | 18.7 | 9.57 | 5.45 | 2.87 | 0.84 | -0.34 |
| $\varepsilon_2 aGST$ | 0 | 0 | 1.32 | 5.43 | 14.9 | 18 | 17.1 | 14.9 | 13.3 | 11.5 |